\documentclass[twocolumn]{aastex631}

\usepackage{amsmath}
\usepackage{amssymb}
\usepackage{xspace}
\usepackage[normalem]{ulem}
\usepackage{soul}
\usepackage{hyperref,url}

\newcommand{\ktwo}{{\it K2}\xspace}
\newcommand{\kep}{\textit{Kepler}\xspace}

\newcommand{\system}{K2-139\xspace}
\newcommand{\planet}{K2-139\,b\xspace}

\newcommand{\feh}{\mbox{[Fe/H]}\xspace}
\newcommand{\teff}{\mbox{$T_{\rm eff}$}\xspace}
\newcommand{\logg}{\mbox{$\log g_{\star}$}\xspace}
\newcommand{\vsini}{\mbox{$v \sin i_{\star}$}\xspace}
\newcommand{\kms}{\mbox{km\,s$^{-1}$}\xspace}
\newcommand{\ms}{\mbox{m\,s$^{-1}$}\xspace}
\newcommand{\mplanet}{\mbox{$M_{\rm p}$}\xspace}
\newcommand{\rplanet}{\mbox{$R_{\rm p}$}\xspace}
\newcommand{\mjup}{\mbox{$\mathrm{M_{\rm Jup}}$}\xspace}
\newcommand{\rjup}{\mbox{$\mathrm{R_{\rm Jup}}$}\xspace}
\newcommand{\mstar}{\mbox{$M_{\star}$}\xspace}
\newcommand{\rstar}{\mbox{$R_{\star}$}\xspace}
\newcommand{\densstar}{\mbox{$\rho_{\star}$}\xspace}
\newcommand{\msol}{\mbox{$\mathrm{M_\odot}$}\xspace}
\newcommand{\rsol}{\mbox{$\mathrm{R_\odot}$}\xspace}
\newcommand{\porb}{\mbox{$P_\mathrm{orb}$}\xspace}
\newcommand{\SIdens}{\mbox{$\mathrm{kg\,m}^{-3}$}\xspace}

\newcommand{\tlcm}{{\sc TLCM}\xspace}
\newcommand{\ariadne}{{\sc ariadne}\xspace}

\newcommand{\PERIODBAR}{$28.38236 \pm 0.00026$}
\newcommand{\EPOCHBAR}{$7325.81714 \pm 0.00033$}
\newcommand{\ECCBAR}{$0.12\substack{+0.12 \\ -0.08}$}
\newcommand{\OMBAR}{$124\substack{+175 \\ -79}$}
\newcommand{\DURATIONBAR}{$ 4.89\substack{+0.08 \\ -0.22}$}
\newcommand{\KAmpBAR}{$27.7\substack{+6.0 \\ -5.3}$}
\newcommand{\ARSTARBAR}{$44.8 \substack{+4.7 \\ -6.7}$}
\newcommand{\RPRSBAR}{$0.0961 \substack{+0.0023 \\ -0.0015}$}
\newcommand{\IMPACTBAR}{$0.30 \substack{+0.21 \\ -0.19}$}
\newcommand{\INCLINATIONBAR}{$89.62 \substack{+0.25 \\ -0.36}$}
\newcommand{\UPLUSBAR}{$0.53 \pm 0.25$} 
\newcommand{\UMINUSBAR}{$0.47 \pm 0.42$} 

\newcommand{\SecondaryTeffBAR}{$565 \substack{+48 \\ -32}$}
\newcommand{\SecondaryMassBAR}{$0.387 \substack{+0.083 \\ -0.075}$}
\newcommand{\SecondaryRadiusBAR}{$0.808 \substack{+0.034 \\ -0.033}$}
\newcommand{\SecondaryDensityBAR}{$910 \substack{+240 \\ -200}$}
\newcommand{\SMABAR}{$0.179 \substack{+0.021 \\ -0.027}$}

\newcommand{\PERIODCH}{$28.38282 \pm 0.00016$}
\newcommand{\EPOCHCH}{$9057.1673 \pm 0.0014$}
\newcommand{\ARSTARCH}{$47.6 \substack{+1.1 \\ -2.1}$}
\newcommand{\RPRSCH}{$0.0955 \substack{+0.0028 \\ -0.0026}$}
\newcommand{\IMPACTCH}{$0.16 \substack{+0.18 \\ -0.12}$}
\newcommand{\UPLUSCH}{$0.64 \pm 0.17$}
\newcommand{\UMINUSCH}{$0.46 \pm 0.37$}

\newcommand{\SMACH}{$0.1747 \pm 0.0016$}
\newcommand{\INCLINATIONCH}{$89.80 \substack{+0.14 \\ -0.23}$}
\newcommand{\DURATIONCH}{$4.912 \substack{+0.098 \\ -0.078} $}
\newcommand{\SecondaryMassCH}{$0.382 \pm 0.078$}
\newcommand{\SecondaryRadiusCH}{$0.802 \substack{+0.029 \\ -0.027}$}
\newcommand{\SecondaryDensityCH}{$912 \substack{+222 \\ -206}$}
\newcommand{\SecondaryTeffCH}{$538 \substack{+13 \\ -9}$}

\newcommand{\PERIODJ}{$28.382796 \pm 0.000042$}
\newcommand{\EPOCHJ}{$9057.16689 \pm 0.00083$}
\newcommand{\ARSTARJ}{$48.0 \substack{+0.9 \\ -1.7}$}
\newcommand{\RPRSJ}{$0.0957 \pm 0.0021$}
\newcommand{\IMPACTJ}{$0.15 \substack{+0.15 \\ -0.11}$}
\newcommand{\UPLUSJ}{$0.62 \pm 0.15$}
\newcommand{\UMINUSJ}{$0.46 \pm 0.37$}

\newcommand{\SMAJ}{$0.1747 \pm 0.0016$}
\newcommand{\INCLINATIONJ}{$89.82 \substack{+0.12 \\ -0.19}$}
\newcommand{\DURATIONJ}{$4.891 \substack{+0.084 \\ -0.067} $}
\newcommand{\SecondaryMassJ}{$0.382 \pm 0.078$}
\newcommand{\SecondaryRadiusJ}{$0.803 \substack{+0.024 \\ -0.023}$}
\newcommand{\SecondaryDensityJ}{$912 \substack{+212 \\ -198}$}
\newcommand{\SecondaryTeffJ}{$536 \substack{+11 \\ -8}$}

\shorttitle{CHEOPS Observations of K2-139}
\shortauthors{A.~M.~S.~Smith \& Sz.~Csizmadia}
\graphicspath{{./}{figures/}}

\begin{document}

\title{A CHEOPS Search for Massive, Long-Period Companions to the Warm Jupiter K2-139\,b}

\author[0000-0002-2386-4341]{Alexis M. S. Smith}\email{alexis.smith@dlr.de}
\affiliation{Department of Extrasolar Planets and Atmospheres, \\
Institute of Planetary Research, \\
German Aerospace Center (DLR),\\ Rutherfordstra\ss e 2 \\ 12489 Berlin, Germany}

\author[0000-0001-6803-9698]{Szil\'ard Csizmadia}
\affiliation{Department of Extrasolar Planets and Atmospheres, \\
Institute of Planetary Research, \\
German Aerospace Center (DLR),\\ Rutherfordstra\ss e 2 \\ 12489 Berlin, Germany}



\begin{abstract}

\planet is a warm Jupiter with an orbital period of 28.4 d, but only three transits of this system have previously been observed, in the long-cadence mode of \ktwo, limiting the precision with which the orbital period can be determined, and future transits predicted. We report photometric observations of four transits of \planet with ESA's CHaracterising ExOPlanet Satellite (CHEOPS), conducted with the goal of measuring the orbital obliquity via spot-crossing events. We jointly fit these CHEOPS data alongside the three previously-published transits from the \ktwo mission, considerably increasing the precision of the ephemeris of \planet. The transit times for this system can now be predicted for the next decade with a $1 \sigma$ precision less than 10 minutes, compared to over one hour previously, allowing the efficient scheduling of observations with {\it Ariel}. We detect no significant deviation from a linear ephemeris, allowing us to exclude the presence of a massive outer planet orbiting with a period less than 150~d, or a brown dwarf with a period less than one year. We also determine the scaled semi-major axis, the impact parameter, and the stellar limb-darkening with improved precision. This is driven by the shorter cadence of the CHEOPS observations compared to that of \ktwo, and validates the sub-exposure technique used for analysing long-cadence photometry. Finally, we note that the stellar spot configuration has changed from the epoch of the \ktwo observations; unlike the \ktwo transits, we detect no evidence of spot-crossing events in the CHEOPS data.

\end{abstract}



\section{Introduction}

The atmospheres of many transiting exoplanets will be probed by {\it Ariel}, ESA's M4 mission, which is due to launch in 2029 \citep{Ariel}. Among the list of potential targets for {\it Ariel} is \planet \citep{Ariel_targets}. Maintaining accurate ephemerides of these targets is vital for the efficient operation of the mission. \planet is one of the targets of interest of the ExoClock Project\footnote{\href{https://www.exoclock.space}{https://www.exoclock.space}} \citep{exo_clock,exo_clock2}, which aims to monitor the {\it Ariel} targets and improve their ephemerides, and is listed as a `high' priority target as of 2022 March.

The issue of decaying ephemerides has also been studied recently from the point-of-view of NASA's Transiting Exoplanet Survey Satellite (TESS; \citealt{TESS}), which is expected to supply many of the bright targets for atmospheric characterization with {\it Ariel} and the James Webb Space Telescope (JWST; \citealt{JWST}). \cite{Dragomir20} found that if no further transits are observed, then around 80 per cent of the best exoplanetary targets for JWST (as selected by \citealt{Kempton18}) would have transit timing uncertainties greater than 30~minutes by the time of the earliest possible JWST observations.

Many systems will be (re-)observed by TESS, which has completed its initial survey of the southern (Cycle 1) and northern (Cycle 2) skies, and is currently re-observing many of the Cycle 1 and 2 sectors. \cite{klagyivik_corot_tess}, for instance, recently updated the ephemerides of several transiting exoplanetary systems originally discovered by CoRoT \citep{Moutou13,corot_final}. However, according to the Web TESS Viewing Tool\footnote{\href{https://heasarc.gsfc.nasa.gov/cgi-bin/tess/webtess/wtv.py}{https://heasarc.gsfc.nasa.gov/cgi-bin/tess/webtess/wtv.py}}, no observations of \system are planned within the first five years of TESS operations (up to and including Sector 69). Many systems can also be observed from the ground, with relatively small telescopes (e.g. \citealt{orbyts2, orbyts1, orbyts3}), however longer-period systems such as \planet are challenging targets for ground-based observations given their less frequent transits and longer transit durations.

\planet was discovered during the re-purposed \kep satellite's \ktwo mission \citep{K2}. Photometry from \ktwo's Campaign 7 revealed three transits of \planet, and the system was subsequently observed and confirmed as a true planetary system by  \cite{K2-139}, hereafter `B18'. According to B18, \planet is a warm Jupiter with a radius of $0.808 \pm 0.034$~\rjup orbiting an active K0V star every 28.4~d. Radial velocity (RV) measurements of the star from the FIES, HARPS, and HARPS-N spectrographs allowed the mass of the planet to be determined as $0.387\substack{+0.083\\-0.075}$~\mjup. The 1$\sigma$ uncertainty on the \planet transit time predicted by the ephemeris of B18 reaches 30~minutes by early 2022, the time of the earliest JWST observations, and is well over an hour by the end of the decade, when {\it Ariel} may observe the system. 

The spin-orbit angle or obliquity, i.e. the angle between the axis of stellar rotation and the orbital axis, of warm Jupiters is particularly interesting. Dynamical migration, whether through Kozai-Lidov cycles \citep{Kozai,Lidov,Fabrycky_Tremaine07} or planet -- planet scattering \citep{R+F96, W+M96} is expected to lead to planets on  high-obliquity orbits. Warm Jupiters (unlike the closer-orbiting hot Jupiters\footnote{Throughout this work, we define hot Jupiters as having $\mplanet \geq 0.3$~\mjup and $\porb < 10$~d, and warm Jupiters as planets with $\mplanet \geq 0.3$~\mjup and $10 \mathrm{d} \leq \porb < 100$~d.}) are not thought to experience strong-enough stellar tidal forces to later co-planarize their orbits. This means that the obliquities of warm Jupiters offer a unique insight into migration history.

While the obliquity has been measured for a significant sample of hot Jupiters, mostly via the Rossiter-McLaughlin effect, but also through Doppler tomographic observations, gravity-darkened transits, and by spot-crossing events, this is not the case for the warm Jupiters. According to TEPCat\footnote{\url{https://www.astro.keele.ac.uk/jkt/tepcat/tepcat.html}, accessed 2022 March 4} \citep{SWorth_homo4}, there are only nine obliquity measurements for warm Jupiters, compared to 120 for hot Jupiters. This is a result both of there being fewer known transiting warm Jupiters (49) than hot Jupiters (453), and of the difficulty, noted above, of observing the transits of longer-period planets from the ground.

Anomalies compatible with spot-crossing events were observed in all three transits observed by \ktwo, and B18 note that this could be exploited to measure the obliquity of the system. This technique has been previously applied to the WASP-4 system \citep{spots_wasp4} among others. Here, we present CHEOPS observations of \system with the goal of constraining the obliquity, as well as updating the orbital ephemeris, of \planet.

\section{CHEOPS observations and data reduction}

\begin{table*}
\caption{Log of CHEOPS observations of \system}
\begin{tabular}{llllll}
\hline
Visit no. & Start date  & Duration & No. of & File\_key & Efficiency$^\dagger$ \\
& (UTC) & (h) & data points & & (\%) \\
\hline
1 & 2020 Jun 27 23:29 & 11.52 & 461 & CH\_PR210005\_TG000101\_V0200 & 69.4  \\
2 & 2020 Jul 26 11:38 & 11.52 & 561 & CH\_PR210005\_TG000201\_V0200 & 64.2  \\
3 & 2021 Jun 03 14:53 & 11.51 & 406 & CH\_PR210005\_TG000202\_V0200 & 71.7  \\
4 & 2021 Jul 01 22:33 & 11.46 & 469 & CH\_PR210005\_TG000203\_V0200 & 66.6  \\
\hline
\\
\end{tabular}
\\
$^\dagger$ The efficiency is the fraction of the visit spent collecting data.
\label{tab:obslog}
\end{table*}

\begin{table}
\caption{CHEOPS light curve of \system, produced by the CHEOPS DRP v13, using the OPTIMAL aperture radius. Only a sample is given here; the full table is available electronically from the CDS}
\begin{tabular}{llll}
\hline
$\mathrm{BJD_{TDB}} $  & Flux & Flux  & Spacecraft \\
- 2\,450\,000&&uncertainty& roll angle ($\degr$) \\
\hline
9028.503166 & 0.988975 & 0.000755 & 272.762 \\
9028.503861 & 0.990507 & 0.000744 & 269.035 \\ 
9028.504556 & 0.990557 & 0.000737 & 265.331 \\
\hline
\\
\end{tabular}
\label{tab:rawlc}
\end{table}

The CHaracterising ExOPlanet Satellite (CHEOPS; \citealt{CHEOPS}) is a small (0.32-m diameter) telescope designed for high-precision monoband photometry of individual exoplanetary systems. CHEOPS was launched in 2019 December, and science operations commenced in 2020 April, with several studies (e.g. \citealt{cheops_wasp189, cheops_wasp103}) already published reporting results from the mission. We observed four transits of \planet with CHEOPS, under the Guest Observers' (GO) programme AO-1-005 (PI: Smith). 

The original aim was to observe four consecutive transits, to see if we could constrain the orbital obliquity of the system. Given the stellar rotation period of $17.24 \pm 0.12$~d (B18), and the orbital period of 28.4~d, we would expect around 29 per cent of the stellar hemisphere visible during the first transit to be visible during the second transit, 41 per cent during the third, and 88 per cent during the fourth transit\footnote{These calculations neglect the effects of differential rotation, which is a reasonable approximation, since the transit impact parameter is small, and so if the system is well-aligned, the planet will transit near-equatorial stellar latitudes.}. We could therefore have expected to observe \planet crossing the same spots in multiple transits if the obliquity is close to zero, and the spots have sufficient longevity, which B18 suggests they may.

Due to scheduling constraints, however, we in fact observed two consecutive transits in 2020 June and July, and a second pair of consecutive transits in 2021 June and July. Each `visit' consisted of seven consecutive orbits of the CHEOPS satellite, or around 11.5 hours. With a $G$-band magnitude of 11.5, \system is a relatively faint target for CHEOPS, and therefore an exposure time of 60~s was used. Further details of the observations are given in Table~\ref{tab:obslog}.

\begin{figure*}
\centering
\includegraphics[angle=270, width=18cm]{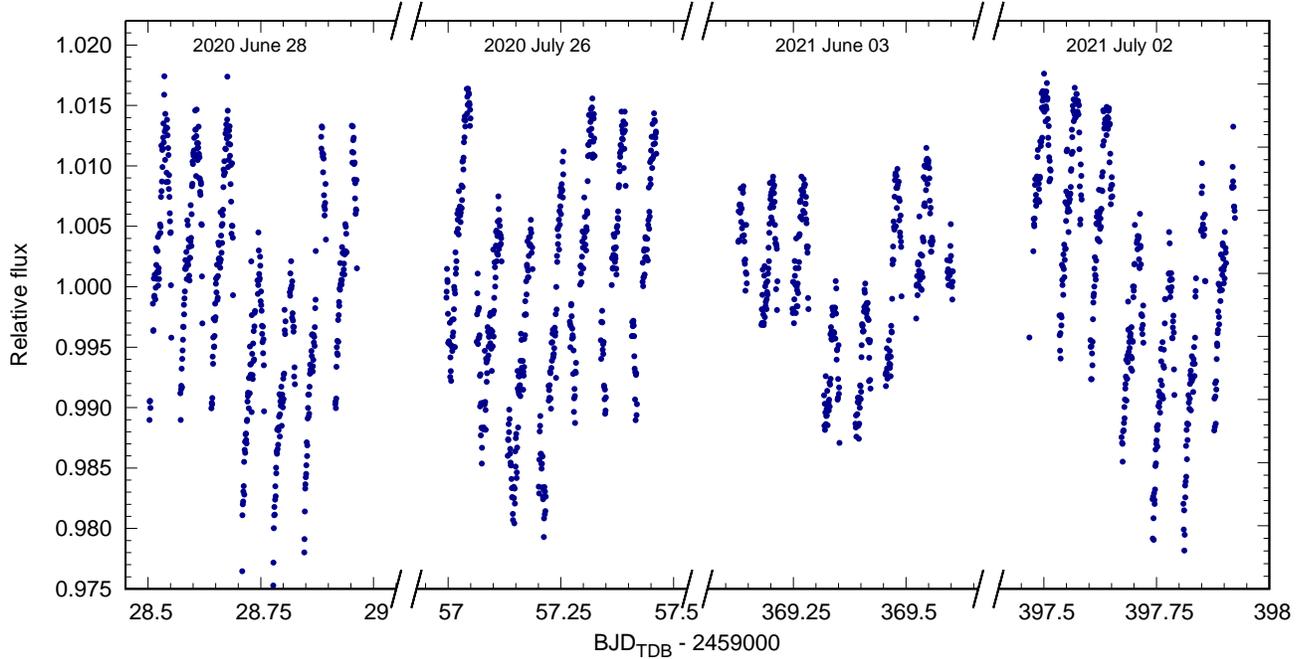} 
\caption{CHEOPS photometry of \system produced by the DRP using the OPTIMAL aperture.
}
\label{fig:raw_phot} 
\end{figure*}

\label{sec:reduction}
The data were reduced using the standard CHEOPS data reduction pipeline (DRP v13; \citealt{cheops_drp}), which produces light curves using four different aperture radii: a `DEFAULT' radius of 25 pixels\footnote{Note that the pixel scale of CHEOPS is approximately 1 pixel $= 1^{\prime \prime}$ \citep{CHEOPS}}, a slightly smaller `RINF' radius of 22 pixels, a larger `RSUP' radius of 30 pixels, and an `OPTIMAL' radius. The size of the latter is determined on a per-visit basis, to account for the differing brightnesses of different targets, and is determined by minimising the noise-to-signal ratio \citep{cheops_drp}. All four visits of \system have the same optimized aperture radius of 15 pixels. We extracted each of these four light curves, corrected for the `ramp' effect, using {\sc pycheops} \citep{pycheops}. We tested each light curve, finding as expected that the light curves produced using the OPTIMAL radius give the highest signal-to-noise; this light curve is shown in Fig.~\ref{fig:raw_phot}, and in Table~\ref{tab:rawlc}.

\section{Stellar characterization}

The previous characterization of \system relied on a co-added FIES spectrum (B18). Here, we update the stellar characterization by fitting the spectral energy distribution (SED) of \system using \ariadne\footnote{\url{{https://github.com/jvines/astroARIADNE}}} \citep{ariadne}. The {\tt {Phoenix~v2}} \citep{Husser2013}, {\tt {BtSettl}} \citep{Allard2012}, \citet{Castelli2003}, and \citet{Kurucz1993} stellar atmospheric model grids were fitted to catalog photometry (Table~\ref{tab:mags}), constrained by the {\it Gaia} parallax. \ariadne uses multiple model grids in order to account for the systematic errors arising from differences between different stellar atmosphere models. Gaussian priors from B18's characterization were applied to \teff and \feh, and the distance was constrained using the Gaia~DR2 parallax \citep{GaiaDR2}. Reddening was accounted for, with $A_V$ limited to the maximum line-of-sight value from the SFD Galactic dust map \citep{Schlegel1998,Schlafly2011}. Bayesian model averaging is employed by \ariadne to derive the uncertainties from the posterior parameter distribution. The distributions for each model are averaged, weighted by the relative probability of each model, leading to smaller uncertainties than those obtained from any single model. \ariadne has previously been used to characterize the primary star in several systems, both eclipsing binaries \citep{Acton_Mdwarf}, and  exoplanetary systems \citep{ngts14,k299_2}.

The stellar parameters resulting from our SED fit are listed in Table~\ref{tab:stellar}, where they are compared to the parameters from B18. Most parameters are in excellent agreement (discrepancies less than  $1 \sigma$), but the stellar age from \ariadne is significantly different to the gyrochronological value of B18. However, we note that the $3\sigma$ limits to the age from \ariadne extend from as young as 43~Myr up to the current age of the Universe; we conclude that the stellar age is poorly-determined. We use the new stellar mass and radius, along with the RV semi-amplitude measured by B18 to determine the mass and radius of \planet (Table~\ref{tab:param}).

\begin{table}
\caption{Stellar parameters of \system.}
\begin{tabular}{cccc}\hline
Parameter & Unit & Spectral value & \ariadne value \\
&& (B18) & (this work)\\
\hline
\teff & K & $5340\pm110$ & $5245\substack{+73\\-33}$ \\
\logg & [cgs] & $4.50\pm0.09$ & $4.53\pm0.05$ \\
\feh & dex & $0.22\pm0.08$ & $0.20\substack{+0.06\\-0.09}$ \\
\vsini & \kms  & $2.8\pm0.6$ & -- \\
Distance & pc  & $152\pm10$ & $156.5\substack{+2.0\\-2.4}$ \\
Age & Gyr   & $1.8\pm0.3$ & $8.9\substack{+1.5\\-1.1}$ \\
$A_V$ & mag   & $0.07\pm0.05$ & $0.05\substack{+0.06\\-0.03}$ \\
\rstar & \rsol   & $0.862\pm0.032$ & $0.863\substack{+0.019\\-0.014}$ \\
\mstar & \msol   & $0.919\pm0.033$ & $0.883\substack{+0.021\\-0.029}$ \\
\hline
\end{tabular}
\label{tab:stellar}
\end{table}

\section{Modelling the photometry}
\subsection{Transit model}

We model the CHEOPS photometry using the Transit Light Curve Modeller (\tlcm; \citealt{tlcm}). The transits are modelled with the following fitted parameters: the planetary orbital period (\porb), the transit epoch ($T_0$), the scaled orbital semi-major axis ($a/ \rstar$), the planet-to-star radius ratio ($\rplanet /\rstar$), the transit impact parameter ($b$), and the limb-darkening coefficients $u_+$ and $u_-$ (which are related to the quadratic coefficients $u_a$ and $u_b$ by $u_+ = u_a + u_b$ and $u_- = u_a - u_b$).

\subsection{Systematics}

In addition to the transit signal, we model the dominant source of systematic noise, which arises from the rotation of the satellite on its orbital period of 98.7 minutes \citep{CHEOPS}. The roll angle of the spacecraft is provided as an additional light curve column by the DRP (see Table~\ref{tab:rawlc}), and is used to model the systematics. Adopting a similar approach to that employed by the {\sc pycheops} software \citep{pycheops}, we model the roll angle dependent systematics of each CHEOPS visit with a function of the form,
\begin{equation}
    f_\mathrm{RA} = \sum_{j=1}^{n_\mathrm{trig}} = \alpha_j\sin{(j\theta)} + \beta_j\cos{(j\theta)}
    \label{eqn:roll}
\end{equation} 
where $\theta$ is the spacecraft roll angle, and the coefficients $\alpha_j$ and $\beta_j$ are fit as MCMC jump parameters within \tlcm. We experimented with different values of $n_\mathrm{trig}$, and adopted $n_\mathrm{trig} = 2$ after determining that higher order terms offer little-to-no improvement in the $\chi^2$ of the model fit, and a higher Bayesian Information Criterion (BIC) value.

After initial modelling as described above, it was apparent that there are additional sources of noise in the light curves, which we attribute to stellar activity. B18 reported variations in the \ktwo light curve with an amplitude of 1 -- 2 per cent, modulated on the stellar rotation period of $17.24 \pm 0.12$~d. We model this as a polynomial function of time for each CHEOPS visit, with the coefficients of the polynomial fitted in \tlcm. We ultimately choose a linear function of time; we determined using the BIC that higher order terms are not justified. Additionally, we modelled the remaining red noise present in the light curves with the wavelet approach of \cite{wavelets}, as implemented in \tlcm \citep{tlcm_red}, fitting for an additional two parameters, the white noise ($\sigma_\mathrm{w}$) and red noise ($\sigma_\mathrm{r}$) levels. This results in us fitting for a total of 33 parameters: seven parameters to describe the transits, and 26 for the various systematics (comprising four roll angle coefficients and two coefficients in time per transit, plus the two wavelet terms).

We also tested an alternative approach to modelling the roll angle dependent systematics. By excluding the data taken during transit, we plotted the out-of-transit flux as a function of roll angle for each of the four CHEOPS visits. We then convolved these data with a Savitzky-Golay filter, and interpolated the resulting function for all values of the roll angle in the whole light curve, subtracting it from the light curve. We then modelled the filtered light curve as above, but with all $\alpha_j$ and $\beta_j$ set to zero. The resulting parameters are in very good agreement with those obtained above, but have slightly smaller uncertainties, as might be expected from an approach where the uncertainties arising from the detrending are not propagated forward to the final fit. We opt for the slightly more conservative error bars resulting from using Eq.~\ref{eqn:roll}, but note that the alternative method has significantly fewer fit parameters, and so the \tlcm runtime is therefore considerably shorter.

\subsection{Joint fit with \ktwo data}

\begin{figure}
\centering
\includegraphics[width=8cm]{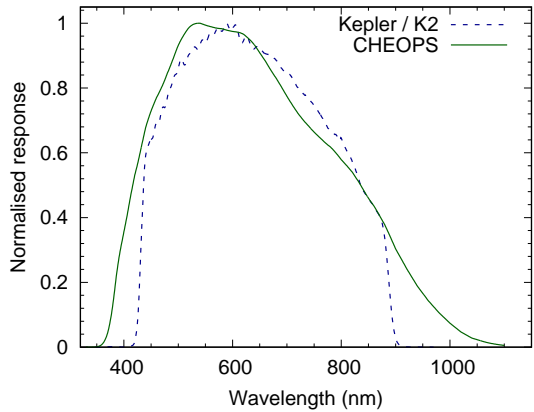} 
\caption{Comparison of \kep/\ktwo (dashed line) and CHEOPS (solid line) instrument response as a function of wavelength. The \kep data are taken from the \kep Science Center, and the CHEOPS data from the ESA website\protect\footnote{\url{https://www.cosmos.esa.int/web/cheops/performances-bandpass}}.
}
\label{fig:bandpass} 
\end{figure}

After modelling the CHEOPS data alone (the results of which are displayed in the middle column of Table~\ref{tab:param}, alongside the corresponding values from B18), we also performed a joint modelling of the CHEOPS and \ktwo photometry. Because the effective exposure time of the \ktwo data is relatively long (1800~s), we subdivide each exposure into five (as was done in e.g. \citealt{k2-137}). We opt not to fit for separate limb-darkening coefficients in the two bandpasses, which is justified because (i) the bandpasses are overlapping and very similar (Fig.~\ref{fig:bandpass}), and (ii) the sparsely-sampled \ktwo data do not enable strong constraints to be placed on the limb-darkening coefficients. 

Table~\ref{tab:limb} lists the limb-darkening coefficients obtained from our various fits, as well as theoretically calculated coefficients for the \kep and CHEOPS bandpasses, computed using the ATLAS stellar models \citep{claret_kepler, claret_cheops}. As expected, given the very similar bandpasses, the differences between the theoretical values for \kep and CHEOPS are small, and much smaller than the uncertainties on our fitted coefficients. The theoretical values also lie well within the 1$\sigma$ uncertainties on our fitted coefficients.

\begin{figure*}
\centering
\includegraphics[width=14.25cm]{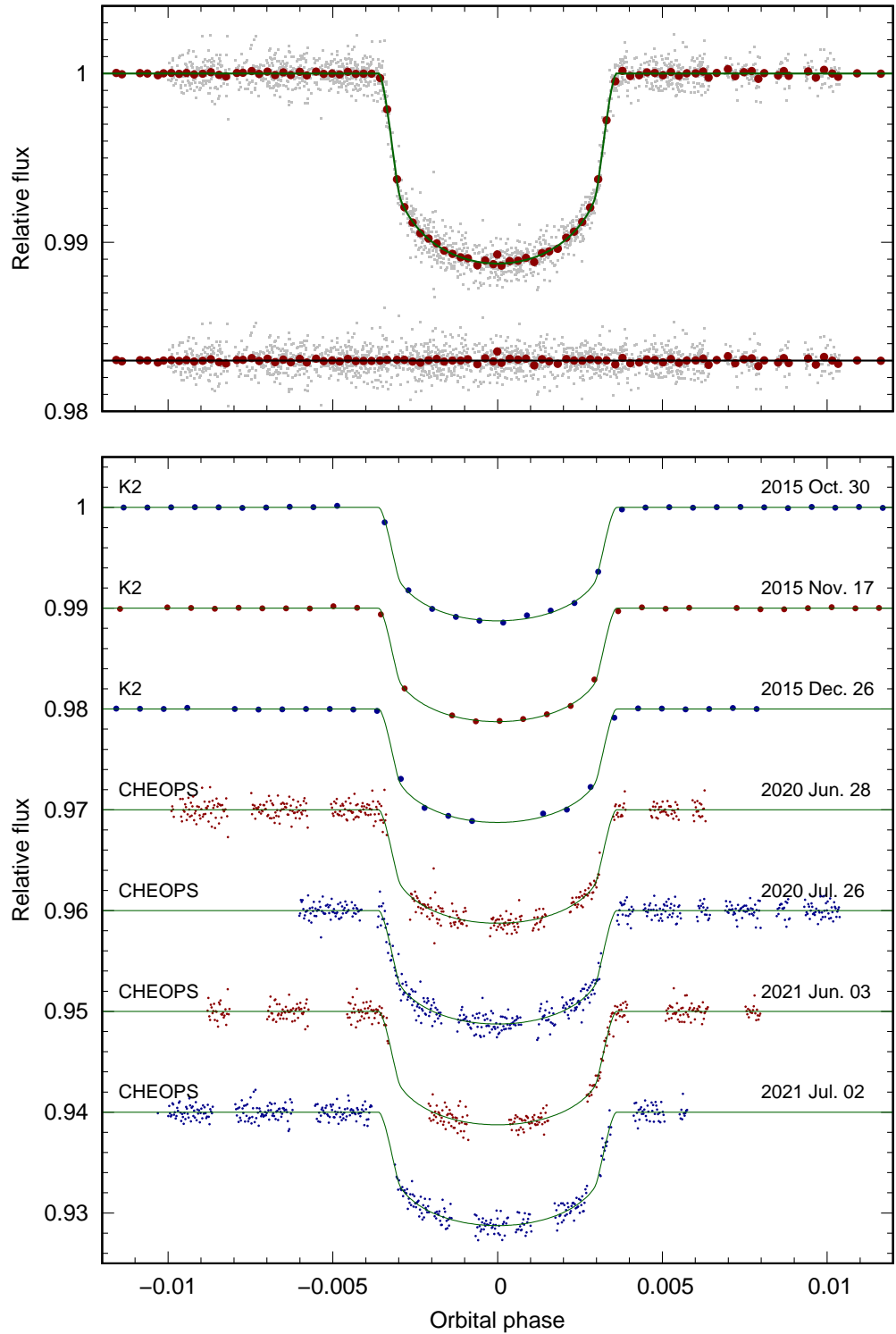} 
\caption{Transit photometry of \system. The top panel shows the combined \ktwo and CHEOPS light curve, phase folded and with all the systematics and the red-noise wavelets subtracted. The original data points are plotted as small grey points, and the larger red circles are the same data binned in phase, with a bin width corresponding to 10 minutes. Below are the residuals to our best-fitting model. The lower panel shows the individual \ktwo and CHEOPS light curves, labelled with the corresponding instrument name and the UT date of mid-transit, and vertically offset for clarity. In all cases, our best-fitting model is shown as a solid green line.
}
\label{fig:fit_phot} 
\end{figure*}

A fit with separate limb-darkening parameters for the two instruments results in coefficients for \ktwo which are compatible with those obtained jointly, but less-precisely determined. Finally, the resulting transit parameters are not affected by our choice to fit for only a single set of limb-darkening coefficients, with the differences between the two parameter sets less than $1\sigma$. The resulting parameters from the joint fit using a common set of limb-darkening coefficients are shown in the final column of Table~\ref{tab:param}, and the photometry is plotted alongside the best-fitting model in Fig.~\ref{fig:fit_phot}.

\subsection{Spot-crossing signal}

Unlike the \ktwo transits, we see no evidence of a spot-crossing signal in any of the CHEOPS transits. In order to be certain that we didn't miss any such signal, we also looked at the light curves and their residuals produced using each of the four photometric apertures (Sect.~\ref{sec:reduction}), fitted both with and without the wavelet red noise model. We also verified that a spot-crossing signal of the type seen in the \ktwo data would not be removed by our modelling of the systematic noise in the CHEOPS light curves. We did this by modelling the \ktwo spot-crossing signal with a toy model, and injecting this signal into the CHEOPS light curves, before fitting them with the procedure described above. This injected signal has an amplitude of 800~ppm, and a duration of 120 minutes. For comparison, the longest gap in the CHEOPS light curves is around 50 minutes, with gaps of 20 -- 30 minutes more typical. The signal was still visible in the CHEOPS light curves, even after fitting them with the wavelet model. In common with B18, we find that excluding the \ktwo points affected by the spot crossings results in a change to the transit parameters which is much smaller than the 1~$\sigma$ uncertainties on those parameters. We therefore opt not to exclude these points from the final fit.

\subsection{Transit timing}
\label{sec:modelttv}
\begin{figure}
\centering
\includegraphics[angle=270,width=8cm]{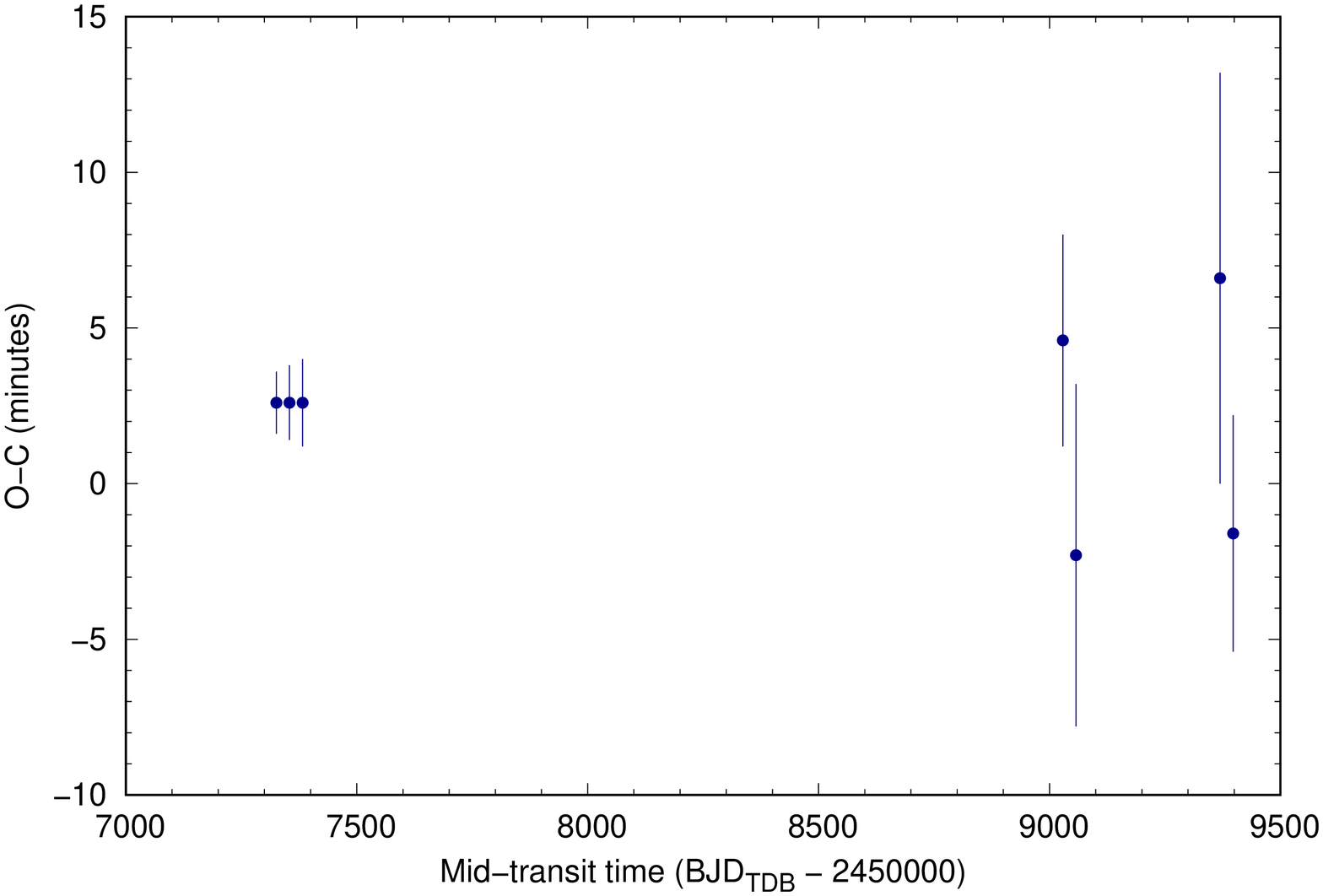} 
\caption{Transit times of \planet. The differences between the observed transit times and our newly-derived ephemeris ($O-C$) are plotted against the mid-transit time (See Table~\ref{tab:ttv}).
}
\label{fig:OC} 
\end{figure}

To measure the time of mid-transit for each of the observed transits of \planet, we re-fit each transit light curve individually, with \porb fixed to the value of our joint fit, and with priors corresponding to the posterior distribution of our joint fit placed on the other transit parameters. These transit times and their uncertainties are listed in Table~\ref{tab:ttv}, where they are also compared to the ephemeris reported by B18, and the ephemeris determined from our joint fit. Although the four CHEOPS transits show significant timing deviation from the B18 ephemeris, this is due to the insufficient precision of the period determined by B18, from just three sparsely-sampled transits. Our newly-determined \porb is six times more precise, and with this ephemeris, there is no evidence for any significant TTVs -- the measured deviations of 2 -- 7 minutes from our ephemeris are comparable to the timing precision of 1 -- 7 minutes on the individual transits (Fig.~\ref{fig:OC}).

\subsection{Orbital eccentricity and stellar density}
\label{sec:ecc}

Alongside the \ktwo photometry, B18 fitted 19 RV measurements of \system, made with three different telescopes / spectrographs. They determined that a simple Keplerian model was a poor fit to the RVs, which are affected by stellar activity. To model this activity, they fit two additional sine curves to the data, at the stellar rotation period and its first harmonic. This fit resulted in a small orbital eccentricity, $e=$\ECCBAR, and a poorly-determined argument of periastron, $\omega =$ \OMBAR. Since the B18 eccentricity value is consistent with zero at the $2 \sigma$ level, and relatively little information about orbital eccentricity is conveyed by transit light curves, we opt to impose a circular orbit solution, by fixing $e=0$.

The stellar density calculated from the mass and radius produced by \ariadne is $\densstar = 1937 \pm 124$~\SIdens, whereas the density from our fit to the \ktwo and CHEOPS transits is $\densstar = 2593 \pm 211$~\SIdens, which differ by around $2.7\sigma$. We therefore tried fitting the light curves again, this time fixing $e$ and $\omega$ to the best-fitting values of B18. All of the resulting parameters and uncertainties are virtually identical to those from the circular orbit fit (with agreements well within $1 \sigma$), except we find $a/ \rstar = 43.3 \substack{+0.9 \\-1.9}$, which is discrepant from the circular value by $2.4\sigma$. The stellar density calculated with this value of $a/ \rstar$ is $\densstar = 1905 \pm 183$~\SIdens, which is in excellent agreement with the \ariadne value. This suggests that the orbit of \planet may indeed be slightly eccentric. We propose that a new RV study of \system is needed to resolve the question of \planet's orbital eccentricity.

\begin{table*}
\begin{center}
\caption{Parameters from light curve and RV data analysis.\label{tab:param}}
\begin{tabular}{lccc}
\hline
\noalign{\smallskip}
Parameter & B18 & CHEOPS & CHEOPS \& \ktwo\\
\noalign{\smallskip}
\hline
\noalign{\smallskip}
Fitted parameters: &&&\\ 
\hline
\noalign{\smallskip}
Orbital period \porb / d  &              \PERIODBAR  & \PERIODCH &\PERIODJ\\
Transit epoch $T_0$ / BJD$_\mathrm{TDB}-2450000$  &              \EPOCHBAR   &\EPOCHCH & \EPOCHJ\\
Scaled semi-major axis $a/ \rstar$   &      \ARSTARBAR &\ARSTARCH&\ARSTARJ\\
Radius ratio $\rplanet/\rstar$   &  \RPRSBAR & \RPRSCH &\RPRSJ\\
Transit impact parameter $b$     &     \IMPACTBAR & \IMPACTCH &\IMPACTJ\\
Limb-darkening coefficient$^\ddag$ $u_+$ &   \UPLUSBAR & \UPLUSCH &\UPLUSJ\\
Limb-darkening coefficient$^\ddag$ $u_-$ &   \UMINUSBAR  & \UMINUSCH &\UMINUSJ\\
Radial velocity semi amplitude $K$ / \ms &  \KAmpBAR  & -- & -- \\
\noalign{\smallskip}
Derived parameters:&&&\\
\hline
\noalign{\smallskip}
Orbital eccentricity $e$ & \ECCBAR & 0.0 (fixed) &0.0 (fixed)\\
Argument of periastron $\omega$ / degrees & \OMBAR & -- & -- \\
Semi-major axis $a$ / au &    \SMABAR & \SMACH &\SMAJ\\
Orbital inclination angle $i$ / degrees   &    \INCLINATIONBAR    &\INCLINATIONCH& \INCLINATIONJ\\
Transit Duration $T_{14}$ / h        &              \DURATIONBAR                        &\DURATIONCH & \DURATIONJ\\
Planet mass* $M_\mathrm{p} / M_\mathrm{Jup}$                     & \SecondaryMassBAR   &\SecondaryMassCH& \SecondaryMassJ\\
Planet radius $R_\mathrm{p} / R_\mathrm{Jup}$                    & \SecondaryRadiusBAR  &\SecondaryRadiusCH& \SecondaryRadiusJ \\
Planet mean density / kg\,m$^{-3}$          & \SecondaryDensityBAR & \SecondaryDensityCH& \SecondaryDensityJ\\
Planet equilibrium temperature$^\dagger$ $T_\mathrm{p,A=0}$ / K& \SecondaryTeffBAR  &\SecondaryTeffCH& \SecondaryTeffJ\\
\noalign{\smallskip}
\hline
\end{tabular}
\\
* calculated for the CHEOPS and CHEOPS \& \ktwo columns using the $K$ measured by B18, and the stellar mass from Table~\ref{tab:stellar}.
$^{\dagger}$assuming zero albedo, and isotropic heat redistribution.\\
$^{\ddag}$ B18 fit for $q_1$ and $q_2$, converted here to $u_+$ and $u_-$ for ease of comparison.
\end{center}
\end{table*}

\begin{table}
\caption{Fitted times of mid-transit for individual transits of \planet, their uncertainties (in days and in minutes), and the deviations (O-C) from the ephemerides presented in B18 and our updated ephemeris.}
\begin{tabular}{cccccc}\hline
$E$ &$T_{\rm c} - 2\ 450\ 000$ & $\sigma_{T_{\rm c}}$ & $\sigma_{T_{\rm c}}$ & \multicolumn{2}{c}{(O-C) / min} \\
(B18) &$\mathrm{BJD_{TDB}}$ & d & min & B18 & new $P$ \\
\hline
0  & 7325.8181  &  0.0007  &  1.0  &  1.4   &  2.6 \\ 
1  & 7354.2010  &  0.0008  &  1.2  &  2.1   &  2.6 \\ 
2  & 7382.5837  &  0.0010  &  1.4  &  2.7   &  2.6 \\ 
60 & 9028.7873  &  0.0024  &  3.4  &  41.1  &  4.6 \\ 
61 & 9057.1653  &  0.0038  &  5.5  &  34.9  & -2.3 \\ 
72 & 9369.3822  &  0.0046  &  6.6  &  50.6  &  6.6 \\ 
73 & 9397.7593  &  0.0026  &  3.8  &  43.1  & -1.6 \\ 
\hline
\\
\end{tabular}
\label{tab:ttv}
\end{table}

\section{Results and discussion}
\subsection{Updated system parameters and ephemeris}

Our combined modelling of the \ktwo and CHEOPS photometry has resulted in better-determined transit parameters, particularly $a/\rstar$ and the limb-darkening coefficients, which benefit from the shorter cadence of the CHEOPS observations. The precision in $\rplanet/\rstar$ is, however, not improved by the CHEOPS data. This validates the approach of modelling the long-cadence \ktwo data using sub-exposures -- this technique is still able to measure $\rplanet/\rstar$ with high precision, while measuring $a/\rstar$ accurately, if somewhat imprecisely.

As expected, our measurement of the orbital period is considerably more precise than that of B18. The 1$\sigma$ uncertainty on the transit time is now around two minutes at the end of 2022, compared to more than 30 minutes previously, and will still be well under 10 minutes at the end of the current decade. Even at the end of the subsequent decade, the uncertainty will only be around 15 minutes, compared to approximately two hours for the B18 ephemeris. This improvement will allow follow-up observations with e.g. {\it Ariel} to be planned and executed efficiently.

\subsection{Transit timing}

As reported in Sect.~\ref{sec:modelttv} and Table~\ref{tab:ttv}, we find no evidence for TTVs. Our individual transit timings from CHEOPS have relatively large uncertainties, which is caused by poor sampling of the ingress and egress phases of the transits. \cite{cheops_ttv_borsato} measured the transit times of several warm Jupiters with CHEOPS, achieving precisions ranging from around 20~s to about 4~minutes, depending on the brightness of the host star, and the efficiency of the in/egress phases. This range of precision is similar to that achieved by the CoRoT satellite for CoRoT-1b; whose data also suffered from some gaps \citep{corot_1_ttv}. \system is fainter than all of the targets studied by \cite{cheops_ttv_borsato}, and none of our transits has good coverage of both ingress and egress which is needed for the most-precise transit times. Good coverage of both ingress and egress would lead to transit times with a precision of around 30 -- 60~s, as measured with CHEOPS for WASP-103 \citep{cheops_wasp103} which is slightly fainter than \system.

The presence of TTVs could be indicative of the presence of a companion planet; warm Jupiters are found to have companions more frequently than hot Jupiters \citep{Huang16}. A full exploration of the properties of the external perturbers not excluded by our non-detection of TTVs is beyond the scope of this work. We can, however, place indicative upper limits on the maximum allowed mass of a putative outer planet, $M_\mathrm{c}$, for a given orbital period, $P_\mathrm{c}$, and eccentricity, $e_\mathrm{c}$, of the perturber, using Eq.~2 of \cite{Holman&Murray}\footnote{with the location of the $\pi$ corrected as per \cite{Borkovits11}}:
\begin{equation}
    M_\mathrm{c} = \frac{16\pi}{45}\mstar \frac{\Delta t_\mathrm{max}}{P_\mathrm{orb}} \left(\frac{P_\mathrm{c}}{P_\mathrm{orb}}\right)^2 (1-e_\mathrm{c})^3
,
\end{equation}
where $\Delta t_\mathrm{max}$ is the maximum allowed TTV amplitude. We adopt $\Delta t_\mathrm{max} = 15$~minutes, which is approximately three times the mean uncertainty on the timing of the CHEOPS transits.

\begin{figure}
\centering
\includegraphics[angle=270, width=8.5cm]{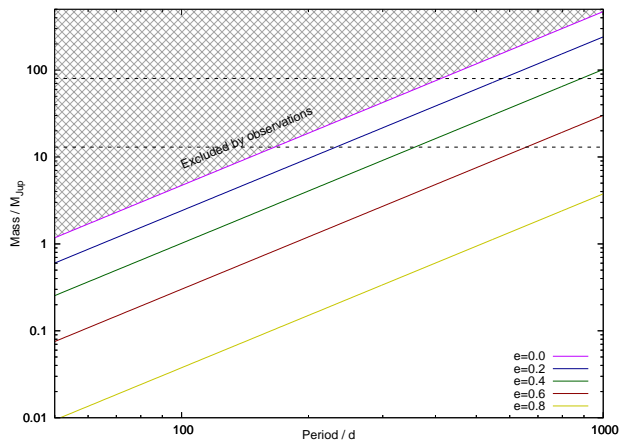} 
\caption{Maximum allowed mass of a perturbing outer body as a function of its orbital period, for different orbital eccentricities of the perturber (solid colored lines). The approximate masses delimiting massive planets from brown dwarfs and brown dwarfs from low-mass stars are indicated with dashed black lines, and the region of parameter space excluded by our TTV observations is shaded.
}
\label{fig:perturber} 
\end{figure}

In Fig.~\ref{fig:perturber} we plot the maximum allowed perturber mass as a function of orbital period for several values of $e_\mathrm{c}$. Our constraints are not particularly strong, but we are able to exclude the presence of a massive planet orbiting with a period less than around 150 days or a brown dwarf with $P_\mathrm{c} \lessapprox 1$~yr. Much of the parameter space, however, is not excluded; for instance the outer massive planet (or brown dwarf) in the K2-99 system, which is clearly visible in the RV measurements of the system \citep{K299,k299_2}, is only expected to induce TTVs for the inner warm Jupiter with an amplitude of just a few minutes. We would not have detected such a signal in our observations of \system, nor would we have detected the outer brown dwarf in the CoRoT-20 system \citep{corot20_deleuil, corot20_rey}.

\subsection{Starspots}

We conclude that there is no evidence for any spot-crossing event in the CHEOPS light curves, which suggests that there were no sufficiently large starspots along the transit chord at the time of these observations. This could be due to chance, or due to the reduced chromospheric activity at a quieter phase of \system's stellar cycle.

Unfortunately this means that we are unable to place any constraint on the obliquity of \planet. Measuring this quantity for warm Jupiters remains very interesting, because the orbital distance of these systems means that their orbits are unlikely to be significantly affected by stellar tidal forces acting to circularize and co-planarize them. This means that an obliquity measurement for \planet would represent the primordial obliquity, following whatever migration processes led the planet to its current orbit. Since different migration processes are expected to result in orbits with different obliquities, measuring this parameter for a sample of warm Jupiters would offer significant insight into migration history. Measuring the obliquity of \planet may be possible via the Rossiter-McLaughlin effect, but any ground-based transit observation of this system is challenging given the lengthy orbital period and transit duration (there is at most one opportunity per year to observe a complete transit of \planet from a given observing site).

\begin{acknowledgments}
CHEOPS is an ESA mission in partnership with Switzerland with important contributions to the payload and the ground segment from Austria, Belgium, France, Germany, Hungary, Italy, Portugal, Spain, Sweden and the United Kingdom.

This work has made use of data from the European Space Agency (ESA) mission
{\it Gaia} (\url{https://www.cosmos.esa.int/gaia}), processed by the {\it Gaia}
Data Processing and Analysis Consortium (DPAC,
\url{https://www.cosmos.esa.int/web/gaia/dpac/consortium}). Funding for the DPAC
has been provided by national institutions, in particular the institutions
participating in the {\it Gaia} Multilateral Agreement.

This research has made use of the NASA Exoplanet Archive, which is operated by the California Institute of Technology, under contract with the National Aeronautics and Space Administration under the Exoplanet Exploration Program.

This publication makes use of The Data \& Analysis Center for Exoplanets (DACE), which is a facility based at the University of Geneva (CH) dedicated to extrasolar planets data visualisation, exchange and analysis. DACE is a platform of the Swiss National Centre of Competence in Research (NCCR) PlanetS, federating the Swiss expertise in Exoplanet research. The DACE platform is available at \url{https://dace.unige.ch}. \\ 
Sz.Cs. is supported by Deutsche Forschungsgemeinschaft Research Unit 2440: ’Matter Under Planetary Interior Conditions: High Pressure Planetary and Plasma Physics’.

\end{acknowledgments}

%

\vspace{5mm}
\facilities{CHEOPS, {\it Gaia}}


\software{\tlcm \citep{tlcm},
          astropy \citep{astropy1,astropy2},  
          }



\appendix

\section{Comparison of limb-darkening coefficients for \system}

\begin{table*}
\caption{Limb-darkening coefficients for \system. Values in italics are calculated from coefficients in other forms.}
\begin{tabular}{lcccc}\hline
                                            &$u_a$          &$u_b$          &$u_+$           &$u_-$           \\
\hline
Theoretical \kep \citep{claret_kepler}	&$0.5256	 $  &$0.1860      $	&\itshape0.7116 &\itshape 0.3396\\
Theoretical CHEOPS \citep{claret_cheops}            &$0.5419	 $  &$0.1689      $ &\itshape 0.7108	&\itshape 0.373\\
B18 fit to \ktwo data	                        &\itshape 0.5$\pm$ 0.24	&\itshape 0.03 $\pm$0.24	&\itshape 0.53 $\pm$ 0.25 &\itshape 0.47 $\pm$ 0.42 \\
Our fit to \ktwo data only	                    &\itshape 0.57$\pm$0.17	&\itshape-0.12$\pm$0.17	&$0.45 \pm 0.10$ &$0.69 \pm 0.32 $\\
Our fit to CHEOPS data only	                &\itshape0.55$\pm$0.20	&\itshape0.09 $\pm$0.20	&$0.64 \pm 0.17$ &$0.46 \pm 0.37 $\\
Our fit to \ktwo \& CHEOPS (common l.d.)	    &\itshape0.54$\pm$0.20	&\itshape0.08 $\pm$0.20	&$0.62 \pm 0.15$ &$0.46 \pm 0.37 $\\
Our fit to \ktwo \& CHEOPS (different l.d.): \ktwo		&\itshape0.35$\pm$0.38 &\itshape0.22$\pm$0.38 &$0.56 \pm0.24 $ &$0.13 \pm 0.72 $\\
Our fit to \ktwo \& CHEOPS (different l.d.): CHEOPS		&\itshape0.43$\pm$0.34	&\itshape0.14$ \pm $0.34 &$0.56     \pm 0.22    $ &$0.29      \pm 0.64    $\\
\hline
\end{tabular}
\label{tab:limb}
\end{table*}

\section{Catalog photometry of \system used in the SED fit}

\begin{table}
\caption{Catalog photometry of \system used in SED fit.}
\begin{tabular}{rr}\hline
Band & Magnitude \\
\hline
2MASS $H$           & $9.7680 \pm 0.0220$ \\
2MASS $J$           & $10.1770 \pm 0.0220$ \\
2MASS $K_s$         & $9.6600 \pm 0.0230$ \\
Johnson $V$         & $11.6780 \pm 0.0130$ \\
Johnson $B$         & $12.5090 \pm 0.0150$ \\
Tycho $B$         & $12.6210 \pm 0.2050$ \\
Tycho $V$         & $11.7400 \pm 0.1370$ \\
GaiaDR2 $G$         & $11.4626 \pm 0.0005$ \\
GaiaDR2 $RP$        & $10.8661 \pm 0.0017$ \\
GaiaDR2 $BP$        & $11.9296 \pm 0.0022$ \\
SDSS $g^\prime$              & $12.0400 \pm 0.0050$ \\
SDSS $i^\prime$              & $11.2250 \pm 0.0120$ \\
SDSS $r^\prime$              & $11.4200 \pm 0.0330$ \\
SkyMapper $u$         & $13.8500 \pm 0.0130$ \\
SkyMapper $v$         & $13.4450 \pm 0.0090$ \\
SkyMapper $g$         & $11.8630 \pm 0.0020$ \\
SkyMapper $r$         & $11.4550 \pm 0.0040$ \\
SkyMapper $i$         & $11.1930 \pm 0.0030$ \\
SkyMapper $z$         & $11.1380 \pm 0.0030$ \\
WISE $W1$         & $9.6040 \pm 0.0230$ \\
WISE $W2$         & $9.6770 \pm 0.0200$ \\
GALEX (NUV)           & $18.3790 \pm 0.0610$ \\
TESS                & $10.9139 \pm 0.0062$ \\
\hline
\end{tabular}
\label{tab:mags}
\end{table}


\bibliography{refs2}{}

\begin{thebibliography}{}
\expandafter\ifx\csname natexlab\endcsname\relax\def\natexlab#1{#1}\fi
\providecommand{\url}[1]{\href{#1}{#1}}
\providecommand{\dodoi}[1]{doi:~\href{http://doi.org/#1}{\nolinkurl{#1}}}
\providecommand{\doeprint}[1]{\href{http://ascl.net/#1}{\nolinkurl{http://ascl.net/#1}}}
\providecommand{\doarXiv}[1]{\href{https://arxiv.org/abs/#1}{\nolinkurl{https://arxiv.org/abs/#1}}}

\bibitem[{{Acton} {et~al.}(2020){Acton}, {Goad}, {Casewell}, {Vines},
  {Burleigh}, {Eigm{\"u}ller}, {Nielsen}, {G{\"a}nsicke}, {Bayliss}, {Bouchy},
  {Bryant}, {Gill}, {Gillen}, {G{\"u}nther}, {Jenkins}, {McCormac}, {Moyano},
  {Raynard}, {Tilbrook}, {Udry}, {Watson}, {West}, \&
  {Wheatley}}]{Acton_Mdwarf}
{Acton}, J.~S., {Goad}, M.~R., {Casewell}, S.~L., {et~al.} 2020, \mnras, 498,
  3115, \dodoi{10.1093/mnras/staa2513}

\bibitem[{Allard {et~al.}(2012)Allard, Homeier, \& Freytag}]{Allard2012}
Allard, F., Homeier, D., \& Freytag, B. 2012, Philosophical Transactions of the
  Royal Society A: Mathematical, Physical and Engineering Sciences, 370, 2765,
  \dodoi{10.1098/rsta.2011.0269}

\bibitem[{{Astropy Collaboration} {et~al.}(2013){Astropy Collaboration},
  {Robitaille}, {Tollerud}, {Greenfield}, {Droettboom}, {Bray}, {Aldcroft},
  {Davis}, {Ginsburg}, {Price-Whelan}, {Kerzendorf}, {Conley}, {Crighton},
  {Barbary}, {Muna}, {Ferguson}, {Grollier}, {Parikh}, {Nair}, {Unther},
  {Deil}, {Woillez}, {Conseil}, {Kramer}, {Turner}, {Singer}, {Fox}, {Weaver},
  {Zabalza}, {Edwards}, {Azalee Bostroem}, {Burke}, {Casey}, {Crawford},
  {Dencheva}, {Ely}, {Jenness}, {Labrie}, {Lim}, {Pierfederici}, {Pontzen},
  {Ptak}, {Refsdal}, {Servillat}, \& {Streicher}}]{astropy1}
{Astropy Collaboration}, {Robitaille}, T.~P., {Tollerud}, E.~J., {et~al.} 2013,
  \aap, 558, A33, \dodoi{10.1051/0004-6361/201322068}

\bibitem[{{Barrag{\'a}n} {et~al.}(2018){Barrag{\'a}n}, {Gandolfi}, {Smith},
  {Deeg}, {Fridlund}, {Persson}, {Donati}, {Endl}, {Csizmadia}, {Grziwa},
  {Nespral}, {Hatzes}, {Cochran}, {Fossati}, {Brems}, {Cabrera}, {Cusano},
  {Eigm{\"u}ller}, {Eiroa}, {Erikson}, {Guenther}, {Korth}, {Lorenzo-Oliveira},
  {Mancini}, {P{\"a}tzold}, {Prieto-Arranz}, {Rauer}, {Rebollido}, {Saario}, \&
  {Zakhozhay}}]{K2-139}
{Barrag{\'a}n}, O., {Gandolfi}, D., {Smith}, A.~M.~S., {et~al.} 2018, \mnras,
  475, 1765 (B18), \dodoi{10.1093/mnras/stx3207}

\bibitem[{{Barros} {et~al.}(2022){Barros}, {Akinsanmi}, {Bou{\'e}}, {Smith},
  {Laskar}, {Ulmer-Moll}, {Lillo-Box}, {Queloz}, {Cameron}, {Sousa},
  {Ehrenreich}, {Hooton}, {Bruno}, {Demory}, {Correia}, {Demangeon}, {Wilson},
  {Bonfanti}, {Hoyer}, {Alibert}, {Alonso}, {Escud{\'e}}, {Barbato},
  {B{\'a}rczy}, {Barrado}, {Baumjohann}, {Beck}, {Beck}, {Benz}, {Bergomi},
  {Billot}, {Bonfils}, {Bouchy}, {Brandeker}, {Broeg}, {Cabrera}, {Cessa},
  {Charnoz}, {Damme}, {Davies}, {Deleuil}, {Deline}, {Delrez}, {Erikson},
  {Fortier}, {Fossati}, {Fridlund}, {Gandolfi}, {Mu{\~n}oz}, {Gillon},
  {G{\"u}del}, {Isaak}, {Heng}, {Kiss}, {des Etangs}, {Lendl}, {Lovis},
  {Magrin}, {Nascimbeni}, {Maxted}, {Olofsson}, {Ottensamer}, {Pagano},
  {Pall{\'e}}, {Parviainen}, {Peter}, {Piotto}, {Pollacco}, {Ragazzoni},
  {Rando}, {Rauer}, {Ribas}, {Santos}, {Scandariato}, {S{\'e}gransan}, {Simon},
  {Steller}, {Szab{\'o}}, {Thomas}, {Udry}, {Ulmer}, {Van Grootel}, \&
  {Walton}}]{cheops_wasp103}
{Barros}, S.~C.~C., {Akinsanmi}, B., {Bou{\'e}}, G., {et~al.} 2022, \aap, 657,
  A52, \dodoi{10.1051/0004-6361/202142196}

\bibitem[{{Benz} {et~al.}(2021){Benz}, {Broeg}, {Fortier}, {Rando}, {Beck},
  {Beck}, {Queloz}, {Ehrenreich}, {Maxted}, {Isaak}, {Billot}, {Alibert},
  {Alonso}, {Ant{\'o}nio}, {Asquier}, {Bandy}, {B{\'a}rczy}, {Barrado},
  {Barros}, {Baumjohann}, {Bekkelien}, {Bergomi}, {Biondi}, {Bonfils},
  {Borsato}, {Brandeker}, {Busch}, {Cabrera}, {Cessa}, {Charnoz}, {Chazelas},
  {Collier Cameron}, {Corral Van Damme}, {Cortes}, {Davies}, {Deleuil},
  {Deline}, {Delrez}, {Demangeon}, {Demory}, {Erikson}, {Farinato}, {Fossati},
  {Fridlund}, {Futyan}, {Gandolfi}, {Garcia Munoz}, {Gillon}, {Guterman},
  {Gutierrez}, {Hasiba}, {Heng}, {Hernandez}, {Hoyer}, {Kiss}, {Kovacs},
  {Kuntzer}, {Laskar}, {Lecavelier des Etangs}, {Lendl}, {L{\'o}pez}, {Lora},
  {Lovis}, {L{\"u}ftinger}, {Magrin}, {Malvasio}, {Marafatto}, {Michaelis}, {de
  Miguel}, {Modrego}, {Munari}, {Nascimbeni}, {Olofsson}, {Ottacher},
  {Ottensamer}, {Pagano}, {Palacios}, {Pall{\'e}}, {Peter}, {Piazza}, {Piotto},
  {Pizarro}, {Pollaco}, {Ragazzoni}, {Ratti}, {Rauer}, {Ribas}, {Rieder},
  {Rohlfs}, {Safa}, {Salatti}, {Santos}, {Scandariato}, {S{\'e}gransan},
  {Simon}, {Smith}, {Sordet}, {Sousa}, {Steller}, {Szab{\'o}}, {Szoke},
  {Thomas}, {Tschentscher}, {Udry}, {Van Grootel}, {Viotto}, {Walter},
  {Walton}, {Wildi}, \& {Wolter}}]{CHEOPS}
{Benz}, W., {Broeg}, C., {Fortier}, A., {et~al.} 2021, Experimental Astronomy,
  51, 109, \dodoi{10.1007/s10686-020-09679-4}

\bibitem[{{Borkovits} {et~al.}(2011){Borkovits}, {Csizmadia},
  {Forg{\'a}cs-Dajka}, \& {Heged{\"u}s}}]{Borkovits11}
{Borkovits}, T., {Csizmadia}, S., {Forg{\'a}cs-Dajka}, E., \& {Heged{\"u}s}, T.
  2011, \aap, 528, A53, \dodoi{10.1051/0004-6361/201015867}

\bibitem[{{Borsato} {et~al.}(2021){Borsato}, {Piotto}, {Gandolfi},
  {Nascimbeni}, {Lacedelli}, {Marzari}, {Billot}, {Maxted}, {Sousa}, {Cameron},
  {Bonfanti}, {Wilson}, {Serrano}, {Garai}, {Alibert}, {Alonso}, {Asquier},
  {B{\'a}rczy}, {Bandy}, {Barrado}, {Barros}, {Baumjohann}, {Beck}, {Beck},
  {Benz}, {Bonfils}, {Brandeker}, {Broeg}, {Cabrera}, {Charnoz}, {Csizmadia},
  {Davies}, {Deleuil}, {Delrez}, {Demangeon}, {Demory}, {des Etangs},
  {Ehrenreich}, {Erikson}, {Escud{\'e}}, {Fortier}, {Fossati}, {Fridlund},
  {Gillon}, {Guedel}, {Hasiba}, {Heng}, {Hoyer}, {Isaak}, {Kiss}, {Kopp},
  {Laskar}, {Lendl}, {Lovis}, {Magrin}, {Munari}, {Olofsson}, {Ottensamer},
  {Pagano}, {Pall{\'e}}, {Peter}, {Pollacco}, {Queloz}, {Ragazzoni}, {Rando},
  {Rauer}, {Ribas}, {S{\'e}gransan}, {Santos}, {Scandariato}, {Simon}, {Smith},
  {Steller}, {Szab{\'o}}, {Thomas}, {Udry}, {Van Grootel}, \&
  {Walton}}]{cheops_ttv_borsato}
{Borsato}, L., {Piotto}, G., {Gandolfi}, D., {et~al.} 2021, \mnras, 506, 3810,
  \dodoi{10.1093/mnras/stab1782}

\bibitem[{{Carter} \& {Winn}(2009)}]{wavelets}
{Carter}, J.~A., \& {Winn}, J.~N. 2009, \apj, 704, 51,
  \dodoi{10.1088/0004-637X/704/1/51}

\bibitem[{{Castelli} \& {Kurucz}(2003)}]{Castelli2003}
{Castelli}, F., \& {Kurucz}, R.~L. 2003, in Modelling of Stellar Atmospheres,
  ed. N.~{Piskunov}, W.~W. {Weiss}, \& D.~F. {Gray}, Vol. 210, A20.
\newblock \doarXiv{astro-ph/0405087}

\bibitem[{{Claret}(2021)}]{claret_cheops}
{Claret}, A. 2021, Research Notes of the American Astronomical Society, 5, 13,
  \dodoi{10.3847/2515-5172/abdcb3}

\bibitem[{{Claret} \& {Bloemen}(2011)}]{claret_kepler}
{Claret}, A., \& {Bloemen}, S. 2011, \aap, 529, A75,
  \dodoi{10.1051/0004-6361/201116451}

\bibitem[{{Csizmadia}(2020)}]{tlcm}
{Csizmadia}, S. 2020, \mnras, 496, 4442, \dodoi{10.1093/mnras/staa349}

\bibitem[{{Csizmadia} {et~al.}(2021){Csizmadia}, {Smith}, {Cabrera},
  {Klagyivik}, {Chaushev}, \& {Lam}}]{tlcm_red}
{Csizmadia}, S., {Smith}, A.~M.~S., {Cabrera}, J., {et~al.} 2021, arXiv
  e-prints, arXiv:2108.11822.
\newblock \doarXiv{2108.11822}

\bibitem[{{Csizmadia} {et~al.}(2010){Csizmadia}, {Renner}, {Barge}, {Agol},
  {Aigrain}, {Alonso}, {Almenara}, {Bonomo}, {Bord{\'e}}, {Bouchy}, {Cabrera},
  {Deeg}, {de La Reza}, {Deleuil}, {Dvorak}, {Erikson}, {Guenther}, {Fridlund},
  {Gondoin}, {Guillot}, {Hatzes}, {Jorda}, {Lammer}, {L{\'a}zaro}, {L{\'e}ger},
  {Llebaria}, {Magain}, {Moutou}, {Ollivier}, {P{\"a}tzold}, {Queloz}, {Rauer},
  {Rouan}, {Schneider}, {Wuchterl}, \& {Gandolfi}}]{corot_1_ttv}
{Csizmadia}, S., {Renner}, S., {Barge}, P., {et~al.} 2010, \aap, 510, A94,
  \dodoi{10.1051/0004-6361/200912052}

\bibitem[{{Deleuil} {et~al.}(2012){Deleuil}, {Bonomo}, {Ferraz-Mello},
  {Erikson}, {Bouchy}, {Havel}, {Aigrain}, {Almenara}, {Alonso}, {Auvergne},
  {Baglin}, {Barge}, {Bord{\'e}}, {Bruntt}, {Cabrera}, {Carpano}, {Cavarroc},
  {Csizmadia}, {Damiani}, {Deeg}, {Dvorak}, {Fridlund}, {H{\'e}brard},
  {Gandolfi}, {Gillon}, {Guenther}, {Guillot}, {Hatzes}, {Jorda}, {L{\'e}ger},
  {Lammer}, {Mazeh}, {Moutou}, {Ollivier}, {Ofir}, {Parviainen}, {Queloz},
  {Rauer}, {Rodr{\'\i}guez}, {Rouan}, {Santerne}, {Schneider}, {Tal-Or},
  {Tingley}, {Weingrill}, \& {Wuchterl}}]{corot20_deleuil}
{Deleuil}, M., {Bonomo}, A.~S., {Ferraz-Mello}, S., {et~al.} 2012, \aap, 538,
  A145, \dodoi{10.1051/0004-6361/201117681}

\bibitem[{{Deleuil} {et~al.}(2018){Deleuil}, {Aigrain}, {Moutou}, {Cabrera},
  {Bouchy}, {Deeg}, {Almenara}, {H{\'e}brard}, {Santerne}, {Alonso}, {Bonomo},
  {Bord{\'e}}, {Csizmadia}, {D{\`\i}az}, {Erikson}, {Fridlund}, {Gandolfi},
  {Guenther}, {Guillot}, {Guterman}, {Grziwa}, {Hatzes}, {L{\'e}ger}, {Mazeh},
  {Ofir}, {Ollivier}, {P{\"a}tzold}, {Parviainen}, {Rauer}, {Rouan},
  {Schneider}, {Titz-Weider}, {Tingley}, \& {Weingrill}}]{corot_final}
{Deleuil}, M., {Aigrain}, S., {Moutou}, C., {et~al.} 2018, \aap, 619, A97,
  \dodoi{10.1051/0004-6361/201731068}

\bibitem[{{Dragomir} {et~al.}(2020){Dragomir}, {Harris}, {Pepper}, {Barclay},
  {Villanueva}, {Ricker}, {Vanderspek}, {Latham}, {Seager}, {Winn}, {Jenkins},
  {Ciardi}, {Furesz}, {Henze}, {Mireles}, {Morgan}, {Quintana}, {Ting}, \&
  {Yahalomi}}]{Dragomir20}
{Dragomir}, D., {Harris}, M., {Pepper}, J., {et~al.} 2020, \aj, 159, 219,
  \dodoi{10.3847/1538-3881/ab845d}

\bibitem[{{Edwards} {et~al.}(2019){Edwards}, {Mugnai}, {Tinetti}, {Pascale}, \&
  {Sarkar}}]{Ariel_targets}
{Edwards}, B., {Mugnai}, L., {Tinetti}, G., {Pascale}, E., \& {Sarkar}, S.
  2019, \aj, 157, 242, \dodoi{10.3847/1538-3881/ab1cb9}

\bibitem[{{Edwards} {et~al.}(2020){Edwards}, {Anisman}, {Changeat}, {Morvan},
  {Wright}, {Yip}, {Abdullahi}, {Ali}, {Amofa}, {Antoniou}, {Arzouni},
  {Bradley}, {Campana}, {Chavda}, {Creswell}, {Gazieva}, {Gudgeon-Sidelnikova},
  {Guha}, {Hayden}, {Huda}, {Hussein}, {Ibrahim}, {Ike}, {Jama}, {Joshi}, {Kc},
  {Keenan}, {Kelly-Smith}, {Khan}, {Korodimos}, {Liang}, {Nogueira},
  {Martey-Botchway}, {Masruri}, {Miyamaru}, {Moalin}, {Monteiro}, {Nawrocka},
  {Musa}, {Nelson}, {Ogunjuyigbe}, {Patel}, {Pereira}, {Ramsey}, {Rasoul},
  {Reetsong}, {Saeed}, {Sander}, {Sanetra}, {Tarabe}, {Tareke}, {Tasneem},
  {Teo}, {Uddin}, {Upadhyay}, {Yanakiev}, {Yatingiri}, {Dunn}, {Kokori},
  {Tsiaras}, {Gomez}, {Tinetti}, \& {Tennyson}}]{orbyts2}
{Edwards}, B., {Anisman}, L., {Changeat}, Q., {et~al.} 2020, Research Notes of
  the American Astronomical Society, 4, 109, \dodoi{10.3847/2515-5172/aba42b}

\bibitem[{{Edwards} {et~al.}(2021{\natexlab{a}}){Edwards}, {Changeat}, {Yip},
  {Tsiaras}, {Taylor}, {Akhtar}, {AlDaghir}, {Bhattarai}, {Bhudia}, {Chapagai},
  {Huang}, {Kabir}, {Khag}, {Khaliq}, {Khatri}, {Kneth}, {Kothari}, {Najmudin},
  {Panchalingam}, {Patel}, {Premachandran}, {Qayyum}, {Rana}, {Shaikh}, {Syed},
  {Theti}, {Zaidani}, {Saraf}, {de Mijolla}, {Caines}, {Kokori}, {Rocchetto},
  {Mallonn}, {Bachschmidt}, {Bosch}, {Bretton}, {Chatelain}, {Deldem}, {Di
  Sisto}, {Evans}, {Fern{\'a}ndez-Laj{\'u}s}, {Guerra}, {Grau Horta}, {Kang},
  {Kim}, {Leroy}, {Lomoz}, {de Haro}, {Hentunen}, {Jongen}, {Molina},
  {Montaigut}, {Naves}, {Raetz}, {Sauer}, {Watkins}, {W{\"u}nsche}, {Zibar},
  {Dunn}, {Tessenyi}, {Savini}, {Tinetti}, \& {Tennyson}}]{orbyts1}
{Edwards}, B., {Changeat}, Q., {Yip}, K.~H., {et~al.} 2021{\natexlab{a}},
  \mnras, 504, 5671, \dodoi{10.1093/mnras/staa1245}

\bibitem[{{Edwards} {et~al.}(2021{\natexlab{b}}){Edwards}, {Ho}, {Osborne},
  {Deen}, {Hathorn}, {Johnson}, {Patel}, {Vogireddy}, {Waddon}, {Ahmed},
  {Bham}, {Campbell}, {Chummun}, {Crossley}, {Dunsdon}, {Hayes}, {Malik},
  {Marsden}, {Mayfield}, {Mitchell}, {Prosser}, {Rabrenovic}, {Smith},
  {Thomas}, {Kokori}, {Tsiaras}, {Tessenyi}, {Tinetti}, \&
  {Tennyson}}]{orbyts3}
{Edwards}, B., {Ho}, C. S.~K., {Osborne}, H. L.~M., {et~al.}
  2021{\natexlab{b}}, arXiv e-prints, arXiv:2111.10350.
\newblock \doarXiv{2111.10350}

\bibitem[{{Fabrycky} \& {Tremaine}(2007)}]{Fabrycky_Tremaine07}
{Fabrycky}, D., \& {Tremaine}, S. 2007, \apj, 669, 1298, \dodoi{10.1086/521702}

\bibitem[{{Gaia Collaboration} {et~al.}(2018){Gaia Collaboration}, {Brown},
  {Vallenari}, {Prusti}, {de Bruijne}, {Babusiaux}, {Bailer-Jones}, {Biermann},
  {Evans}, {Eyer}, {Jansen}, {Jordi}, {Klioner}, {Lammers}, {Lindegren},
  {Luri}, {Mignard}, {Panem}, {Pourbaix}, {Randich}, {Sartoretti}, {Siddiqui},
  {Soubiran}, {van Leeuwen}, {Walton}, {Arenou}, {Bastian}, {Cropper},
  {Drimmel}, {Katz}, {Lattanzi}, {Bakker}, {Cacciari}, {Casta{\~n}eda},
  {Chaoul}, {Cheek}, {De Angeli}, {Fabricius}, {Guerra}, {Holl}, {Masana},
  {Messineo}, {Mowlavi}, {Nienartowicz}, {Panuzzo}, {Portell}, {Riello},
  {Seabroke}, {Tanga}, {Th{\'e}venin}, {Gracia-Abril}, {Comoretto},
  {Garcia-Reinaldos}, {Teyssier}, {Altmann}, {Andrae}, {Audard},
  {Bellas-Velidis}, {Benson}, {Berthier}, {Blomme}, {Burgess}, {Busso},
  {Carry}, {Cellino}, {Clementini}, {Clotet}, {Creevey}, {Davidson}, {De
  Ridder}, {Delchambre}, {Dell'Oro}, {Ducourant},
  {Fern{\'a}ndez-Hern{\'a}ndez}, {Fouesneau}, {Fr{\'e}mat}, {Galluccio},
  {Garc{\'\i}a-Torres}, {Gonz{\'a}lez-N{\'u}{\~n}ez}, {Gonz{\'a}lez-Vidal},
  {Gosset}, {Guy}, {Halbwachs}, {Hambly}, {Harrison}, {Hern{\'a}ndez},
  {Hestroffer}, {Hodgkin}, {Hutton}, {Jasniewicz}, {Jean-Antoine-Piccolo},
  {Jordan}, {Korn}, {Krone-Martins}, {Lanzafame}, {Lebzelter}, {L{\"o}ffler},
  {Manteiga}, {Marrese}, {Mart{\'\i}n-Fleitas}, {Moitinho}, {Mora}, {Muinonen},
  {Osinde}, {Pancino}, {Pauwels}, {Petit}, {Recio-Blanco}, {Richards},
  {Rimoldini}, {Robin}, {Sarro}, {Siopis}, {Smith}, {Sozzetti}, {S{\"u}veges},
  {Torra}, {van Reeven}, {Abbas}, {Abreu Aramburu}, {Accart}, {Aerts},
  {Altavilla}, {{\'A}lvarez}, {Alvarez}, {Alves}, {Anderson}, {Andrei},
  {Anglada Varela}, {Antiche}, {Antoja}, {Arcay}, {Astraatmadja}, {Bach},
  {Baker}, {Balaguer-N{\'u}{\~n}ez}, {Balm}, {Barache}, {Barata}, {Barbato},
  {Barblan}, {Barklem}, {Barrado}, {Barros}, {Barstow}, {Bartholom{\'e}
  Mu{\~n}oz}, {Bassilana}, {Becciani}, {Bellazzini}, {Berihuete}, {Bertone},
  {Bianchi}, {Bienaym{\'e}}, {Blanco-Cuaresma}, {Boch}, {Boeche}, {Bombrun},
  {Borrachero}, {Bossini}, {Bouquillon}, {Bourda}, {Bragaglia}, {Bramante},
  {Breddels}, {Bressan}, {Brouillet}, {Br{\"u}semeister}, {Brugaletta},
  {Bucciarelli}, {Burlacu}, {Busonero}, {Butkevich}, {Buzzi}, {Caffau},
  {Cancelliere}, {Cannizzaro}, {Cantat-Gaudin}, {Carballo}, {Carlucci},
  {Carrasco}, {Casamiquela}, {Castellani}, {Castro-Ginard}, {Charlot},
  {Chemin}, {Chiavassa}, {Cocozza}, {Costigan}, {Cowell}, {Crifo}, {Crosta},
  {Crowley}, {Cuypers}, {Dafonte}, {Damerdji}, {Dapergolas}, {David}, {David},
  {de Laverny}, {De Luise}, {De March}, {de Martino}, {de Souza}, {de Torres},
  {Debosscher}, {del Pozo}, {Delbo}, {Delgado}, {Delgado}, {Di Matteo},
  {Diakite}, {Diener}, {Distefano}, {Dolding}, {Drazinos}, {Dur{\'a}n},
  {Edvardsson}, {Enke}, {Eriksson}, {Esquej}, {Eynard Bontemps}, {Fabre},
  {Fabrizio}, {Faigler}, {Falc{\~a}o}, {Farr{\`a}s Casas}, {Federici},
  {Fedorets}, {Fernique}, {Figueras}, {Filippi}, {Findeisen}, {Fonti},
  {Fraile}, {Fraser}, {Fr{\'e}zouls}, {Gai}, {Galleti}, {Garabato},
  {Garc{\'\i}a-Sedano}, {Garofalo}, {Garralda}, {Gavel}, {Gavras}, {Gerssen},
  {Geyer}, {Giacobbe}, {Gilmore}, {Girona}, {Giuffrida}, {Glass}, {Gomes},
  {Granvik}, {Gueguen}, {Guerrier}, {Guiraud}, {Guti{\'e}rrez-S{\'a}nchez},
  {Haigron}, {Hatzidimitriou}, {Hauser}, {Haywood}, {Heiter}, {Helmi}, {Heu},
  {Hilger}, {Hobbs}, {Hofmann}, {Holland}, {Huckle}, {Hypki}, {Icardi},
  {Jan{\ss}en}, {Jevardat de Fombelle}, {Jonker}, {Juh{\'a}sz}, {Julbe},
  {Karampelas}, {Kewley}, {Klar}, {Kochoska}, {Kohley}, {Kolenberg},
  {Kontizas}, {Kontizas}, {Koposov}, {Kordopatis}, {Kostrzewa-Rutkowska},
  {Koubsky}, {Lambert}, {Lanza}, {Lasne}, {Lavigne}, {Le Fustec}, {Le
  Poncin-Lafitte}, {Lebreton}, {Leccia}, {Leclerc}, {Lecoeur-Taibi},
  {Lenhardt}, {Leroux}, {Liao}, {Licata}, {Lindstr{\o}m}, {Lister}, {Livanou},
  {Lobel}, {L{\'o}pez}, {Managau}, {Mann}, {Mantelet}, {Marchal}, {Marchant},
  {Marconi}, {Marinoni}, {Marschalk{\'o}}, {Marshall}, {Martino}, {Marton},
  {Mary}, {Massari}, {Matijevi{\v{c}}}, {Mazeh}, {McMillan}, {Messina},
  {Michalik}, {Millar}, {Molina}, {Molinaro}, {Moln{\'a}r}, {Montegriffo},
  {Mor}, {Morbidelli}, {Morel}, {Morris}, {Mulone}, {Muraveva}, {Musella},
  {Nelemans}, {Nicastro}, {Noval}, {O'Mullane}, {Ord{\'e}novic},
  {Ord{\'o}{\~n}ez-Blanco}, {Osborne}, {Pagani}, {Pagano}, {Pailler},
  {Palacin}, {Palaversa}, {Panahi}, {Pawlak}, {Piersimoni}, {Pineau}, {Plachy},
  {Plum}, {Poggio}, {Poujoulet}, {Pr{\v{s}}a}, {Pulone}, {Racero}, {Ragaini},
  {Rambaux}, {Ramos-Lerate}, {Regibo}, {Reyl{\'e}}, {Riclet}, {Ripepi}, {Riva},
  {Rivard}, {Rixon}, {Roegiers}, {Roelens}, {Romero-G{\'o}mez}, {Rowell},
  {Royer}, {Ruiz-Dern}, {Sadowski}, {Sagrist{\`a} Sell{\'e}s}, {Sahlmann},
  {Salgado}, {Salguero}, {Sanna}, {Santana-Ros}, {Sarasso}, {Savietto},
  {Schultheis}, {Sciacca}, {Segol}, {Segovia}, {S{\'e}gransan}, {Shih},
  {Siltala}, {Silva}, {Smart}, {Smith}, {Solano}, {Solitro}, {Sordo}, {Soria
  Nieto}, {Souchay}, {Spagna}, {Spoto}, {Stampa}, {Steele},
  {Steidelm{\"u}ller}, {Stephenson}, {Stoev}, {Suess}, {Surdej}, {Szabados},
  {Szegedi-Elek}, {Tapiador}, {Taris}, {Tauran}, {Taylor}, {Teixeira},
  {Terrett}, {Teyssand ier}, {Thuillot}, {Titarenko}, {Torra Clotet}, {Turon},
  {Ulla}, {Utrilla}, {Uzzi}, {Vaillant}, {Valentini}, {Valette}, {van Elteren},
  {Van Hemelryck}, {van Leeuwen}, {Vaschetto}, {Vecchiato}, {Veljanoski},
  {Viala}, {Vicente}, {Vogt}, {von Essen}, {Voss}, {Votruba}, {Voutsinas},
  {Walmsley}, {Weiler}, {Wertz}, {Wevers}, {Wyrzykowski}, {Yoldas},
  {{\v{Z}}erjal}, {Ziaeepour}, {Zorec}, {Zschocke}, {Zucker}, {Zurbach}, \&
  {Zwitter}}]{GaiaDR2}
{Gaia Collaboration}, {Brown}, A.~G.~A., {Vallenari}, A., {et~al.} 2018, \aap,
  616, A1, \dodoi{10.1051/0004-6361/201833051}

\bibitem[{{Gardner} {et~al.}(2006){Gardner}, {Mather}, {Clampin}, {Doyon},
  {Greenhouse}, {Hammel}, {Hutchings}, {Jakobsen}, {Lilly}, {Long}, {Lunine},
  {McCaughrean}, {Mountain}, {Nella}, {Rieke}, {Rieke}, {Rix}, {Smith},
  {Sonneborn}, {Stiavelli}, {Stockman}, {Windhorst}, \& {Wright}}]{JWST}
{Gardner}, J.~P., {Mather}, J.~C., {Clampin}, M., {et~al.} 2006, \ssr, 123,
  485, \dodoi{10.1007/s11214-006-8315-7}

\bibitem[{{Holman} \& {Murray}(2005)}]{Holman&Murray}
{Holman}, M.~J., \& {Murray}, N.~W. 2005, Science, 307, 1288,
  \dodoi{10.1126/science.1107822}

\bibitem[{{Howell} {et~al.}(2014){Howell}, {Sobeck}, {Haas}, {Still},
  {Barclay}, {Mullally}, {Troeltzsch}, {Aigrain}, {Bryson}, {Caldwell},
  {Chaplin}, {Cochran}, {Huber}, {Marcy}, {Miglio}, {Najita}, {Smith},
  {Twicken}, \& {Fortney}}]{K2}
{Howell}, S.~B., {Sobeck}, C., {Haas}, M., {et~al.} 2014, \pasp, 126, 398,
  \dodoi{10.1086/676406}

\bibitem[{{Hoyer} {et~al.}(2020){Hoyer}, {Guterman}, {Demangeon}, {Sousa},
  {Deleuil}, {Meunier}, \& {Benz}}]{cheops_drp}
{Hoyer}, S., {Guterman}, P., {Demangeon}, O., {et~al.} 2020, \aap, 635, A24,
  \dodoi{10.1051/0004-6361/201936325}

\bibitem[{{Huang} {et~al.}(2016){Huang}, {Wu}, \& {Triaud}}]{Huang16}
{Huang}, C., {Wu}, Y., \& {Triaud}, A.~H.~M.~J. 2016, \apj, 825, 98,
  \dodoi{10.3847/0004-637X/825/2/98}

\bibitem[{Husser {et~al.}(2013)Husser, {Wende-von Berg}, Dreizler, Homeier,
  Reiners, Barman, \& Hauschildt}]{Husser2013}
Husser, T.-O., {Wende-von Berg}, S., Dreizler, S., {et~al.} 2013, A{\&}A, 553,
  A6, \dodoi{10.1051/0004-6361/201219058}

\bibitem[{{Kempton} {et~al.}(2018){Kempton}, {Bean}, {Louie}, {Deming}, {Koll},
  {Mansfield}, {Christiansen}, {L{\'o}pez-Morales}, {Swain}, {Zellem},
  {Ballard}, {Barclay}, {Barstow}, {Batalha}, {Beatty}, {Berta-Thompson},
  {Birkby}, {Buchhave}, {Charbonneau}, {Cowan}, {Crossfield}, {de Val-Borro},
  {Doyon}, {Dragomir}, {Gaidos}, {Heng}, {Hu}, {Kane}, {Kreidberg}, {Mallonn},
  {Morley}, {Narita}, {Nascimbeni}, {Pall{\'e}}, {Quintana}, {Rauscher},
  {Seager}, {Shkolnik}, {Sing}, {Sozzetti}, {Stassun}, {Valenti}, \& {von
  Essen}}]{Kempton18}
{Kempton}, E. M.~R., {Bean}, J.~L., {Louie}, D.~R., {et~al.} 2018, \pasp, 130,
  114401, \dodoi{10.1088/1538-3873/aadf6f}

\bibitem[{{Klagyivik} {et~al.}(2021){Klagyivik}, {Deeg}, {Csizmadia},
  {Cabrera}, \& {Nowak}}]{klagyivik_corot_tess}
{Klagyivik}, P., {Deeg}, H.~J., {Csizmadia}, S., {Cabrera}, J., \& {Nowak}, G.
  2021, Frontiers in Astronomy and Space Sciences, 8, 210,
  \dodoi{10.3389/fspas.2021.792823}

\bibitem[{{Kokori} {et~al.}(2021){Kokori}, {Tsiaras}, {Edwards}, {Rocchetto},
  {Tinetti}, {W{\"u}nsche}, {Paschalis}, {Agnihotri}, {Bachschmidt}, {Bretton},
  {Caines}, {Cal{\'o}}, {Casali}, {Crow}, {Dawes}, {Deldem},
  {Deligeorgopoulos}, {Dymock}, {Evans}, {Falco}, {Ferratfiat}, {Fowler},
  {Futcher}, {Guerra}, {Hurter}, {Jones}, {Kang}, {Kim}, {Lee}, {Lopresti},
  {Marino}, {Mallonn}, {Mortari}, {Morvan}, {Mugnai}, {Nastasi}, {Perroud},
  {Pereira}, {Phillips}, {Pintr}, {Raetz}, {Regembal}, {Savage}, {Sedita},
  {Sioulas}, {Strikis}, {Thurston}, {Tomacelli}, \& {Tomatis}}]{exo_clock}
{Kokori}, A., {Tsiaras}, A., {Edwards}, B., {et~al.} 2021, Experimental
  Astronomy, \dodoi{10.1007/s10686-020-09696-3}

\bibitem[{{Kokori} {et~al.}(2022){Kokori}, {Tsiaras}, {Edwards}, {Rocchetto},
  {Tinetti}, {Bewersdorff}, {Jongen}, {Lekkas}, {Pantelidou}, {Poultourtzidis},
  {W{\"u}nsche}, {Aggelis}, {Agnihotri}, {Arena}, {Bachschmidt}, {Bennett},
  {Benni}, {Bernacki}, {Besson}, {Betti}, {Biagini}, {Brandebourg}, {Bretton},
  {Brincat}, {Cal{\'o}}, {Campos}, {Casali}, {Ciantini}, {Crow}, {Dauchet},
  {Dawes}, {Deldem}, {Deligeorgopoulos}, {Dymock}, {Eenm{\"a}e}, {Evans},
  {Esseiva}, {Falco}, {Ferratfiat}, {Fowler}, {Futcher}, {Gaitan}, {Horta},
  {Guerra}, {Hurter}, {Jones}, {Kang}, {Kiiskinen}, {Kim}, {Laloum}, {Lee},
  {Lomoz}, {Lopresti}, {Mallonn}, {Mannucci}, {Marino}, {Mario}, {Marquette},
  {Michelet}, {Miller}, {Mollier}, {Molina}, {Montigiani}, {Mortari}, {Morvan},
  {Mugnai}, {Naponiello}, {Nastasi}, {Neito}, {Pace}, {Papadeas}, {Paschalis},
  {Pereira}, {Perroud}, {Phillips}, {Pintr}, {Pioppa}, {Popowicz}, {Raetz},
  {Regembal}, {Rickard}, {Roberts}, {Rousselot}, {Rubia}, {Savage}, {Sedita},
  {Shave-Wall}, {Sioulas}, {{\v{S}}koln{\'\i}k}, {Smith}, {St-Gelais},
  {Stouraitis}, {Strikis}, {Thurston}, {Tomacelli}, {Tomatis}, {Trevan},
  {Valeau}, {Vignes}, {Vora}, {Vra{\v{s}}{\v{t}}{\'a}k}, {Walter}, {Wenzel},
  {Wright}, \& {Z{\'\i}bar}}]{exo_clock2}
---. 2022, \apjs, 258, 40, \dodoi{10.3847/1538-4365/ac3a10}

\bibitem[{{Kozai}(1962)}]{Kozai}
{Kozai}, Y. 1962, \aj, 67, 591, \dodoi{10.1086/108790}

\bibitem[{{Kurucz}(1993)}]{Kurucz1993}
{Kurucz}, R.~L. 1993, VizieR Online Data Catalog, 6039

\bibitem[{{Lendl} {et~al.}(2020){Lendl}, {Csizmadia}, {Deline}, {Fossati},
  {Kitzmann}, {Heng}, {Hoyer}, {Salmon}, {Benz}, {Broeg}, {Ehrenreich},
  {Fortier}, {Queloz}, {Bonfanti}, {Brandeker}, {Collier Cameron}, {Delrez},
  {Garcia Mu{\~n}oz}, {Hooton}, {Maxted}, {Morris}, {Van Grootel}, {Wilson},
  {Alibert}, {Alonso}, {Asquier}, {Bandy}, {B{\'a}rczy}, {Barrado}, {Barros},
  {Baumjohann}, {Beck}, {Beck}, {Bekkelien}, {Bergomi}, {Billot}, {Biondi},
  {Bonfils}, {Bourrier}, {Busch}, {Cabrera}, {Cessa}, {Charnoz}, {Chazelas},
  {Corral Van Damme}, {Davies}, {Deleuil}, {Demangeon}, {Demory}, {Erikson},
  {Farinato}, {Fridlund}, {Futyan}, {Gandolfi}, {Gillon}, {Guterman}, {Hasiba},
  {Hernandez}, {Isaak}, {Kiss}, {Kuntzer}, {Lecavelier des Etangs},
  {L{\"u}ftinger}, {Laskar}, {Lovis}, {Magrin}, {Malvasio}, {Marafatto},
  {Michaelis}, {Munari}, {Nascimbeni}, {Olofsson}, {Ottacher}, {Ottensamer},
  {Pagano}, {Pall{\'e}}, {Peter}, {Piazza}, {Piotto}, {Pollacco}, {Ratti},
  {Rauer}, {Ragazzoni}, {Rando}, {Ribas}, {Rieder}, {Rohlfs}, {Safa}, {Santos},
  {Scandariato}, {S{\'e}gransan}, {Simon}, {Singh}, {Smith}, {Sordet}, {Sousa},
  {Steller}, {Szab{\'o}}, {Thomas}, {Tschentscher}, {Udry}, {Viotto}, {Walter},
  {Walton}, {Wildi}, \& {Wolter}}]{cheops_wasp189}
{Lendl}, M., {Csizmadia}, S., {Deline}, A., {et~al.} 2020, \aap, 643, A94,
  \dodoi{10.1051/0004-6361/202038677}

\bibitem[{{Lidov}(1962)}]{Lidov}
{Lidov}, M.~L. 1962, \planss, 9, 719, \dodoi{10.1016/0032-0633(62)90129-0}

\bibitem[{{Maxted} {et~al.}(2021){Maxted}, {Ehrenreich}, {Wilson}, {Alibert},
  {Collier Cameron}, {Hoyer}, {Sousa}, {Olofsson}, {Bekkelien}, {Deline},
  {Delrez}, {Bonfanti}, {Borsato}, {Alonso}, {Anglada Escud{\'e}}, {Barrado},
  {Barros}, {Baumjohann}, {Beck}, {Beck}, {Benz}, {Billot}, {Biondi},
  {Bonfils}, {Brandeker}, {Broeg}, {B{\'a}rczy}, {Cabrera}, {Charnoz}, {Corral
  Van Damme}, {Csizmadia}, {Davies}, {Deleuil}, {Demangeon}, {Demory},
  {Erikson}, {Flor{\'e}n}, {Fortier}, {Fossati}, {Fridlund}, {Futyan},
  {Gandolfi}, {Gillon}, {Guedel}, {Guterman}, {Heng}, {Isaak}, {Kiss},
  {Laskar}, {Lecavelier des Etangs}, {Lendl}, {Lovis}, {Magrin}, {Nascimbeni},
  {Ottensamer}, {Pagano}, {Pall{\'e}}, {Peter}, {Piotto}, {Pollacco},
  {Pozuelos}, {Queloz}, {Ragazzoni}, {Rando}, {Rauer}, {Reimers}, {Ribas},
  {Santos}, {Scandariato}, {Simon}, {Smith}, {Steller}, {Swayne}, {Szab{\'o}},
  {S{\'e}gransan}, {Thomas}, {Udry}, {Van Grootel}, \& {Walton}}]{pycheops}
{Maxted}, P.~F.~L., {Ehrenreich}, D., {Wilson}, T.~G., {et~al.} 2021, \mnras,
  \dodoi{10.1093/mnras/stab3371}

\bibitem[{{Moutou} {et~al.}(2013){Moutou}, {Deleuil}, {Guillot}, {Baglin},
  {Bord{\'e}}, {Bouchy}, {Cabrera}, {Csizmadia}, \& {Deeg}}]{Moutou13}
{Moutou}, C., {Deleuil}, M., {Guillot}, T., {et~al.} 2013, \icarus, 226, 1625,
  \dodoi{10.1016/j.icarus.2013.03.022}

\bibitem[{{Price-Whelan} {et~al.}(2018){Price-Whelan}, {Sip{\H{o}}cz},
  {G{\"u}nther}, {Lim}, {Crawford}, {Conseil}, {Shupe}, {Craig}, {Dencheva},
  {Ginsburg}, {VanderPlas}, {Bradley}, {P{\'e}rez-Su{\'a}rez}, {de Val-Borro},
  {Paper Contributors}, {Aldcroft}, {Cruz}, {Robitaille}, {Tollerud},
  {Coordination Committee}, {Ardelean}, {Babej}, {Bach}, {Bachetti}, {Bakanov},
  {Bamford}, {Barentsen}, {Barmby}, {Baumbach}, {Berry}, {Biscani}, {Boquien},
  {Bostroem}, {Bouma}, {Brammer}, {Bray}, {Breytenbach}, {Buddelmeijer},
  {Burke}, {Calderone}, {Cano Rodr{\'\i}guez}, {Cara}, {Cardoso}, {Cheedella},
  {Copin}, {Corrales}, {Crichton}, {D{\textquoteright}Avella}, {Deil},
  {Depagne}, {Dietrich}, {Donath}, {Droettboom}, {Earl}, {Erben}, {Fabbro},
  {Ferreira}, {Finethy}, {Fox}, {Garrison}, {Gibbons}, {Goldstein}, {Gommers},
  {Greco}, {Greenfield}, {Groener}, {Grollier}, {Hagen}, {Hirst}, {Homeier},
  {Horton}, {Hosseinzadeh}, {Hu}, {Hunkeler}, {Ivezi{\'c}}, {Jain}, {Jenness},
  {Kanarek}, {Kendrew}, {Kern}, {Kerzendorf}, {Khvalko}, {King}, {Kirkby},
  {Kulkarni}, {Kumar}, {Lee}, {Lenz}, {Littlefair}, {Ma}, {Macleod},
  {Mastropietro}, {McCully}, {Montagnac}, {Morris}, {Mueller}, {Mumford},
  {Muna}, {Murphy}, {Nelson}, {Nguyen}, {Ninan}, {N{\"o}the}, {Ogaz}, {Oh},
  {Parejko}, {Parley}, {Pascual}, {Patil}, {Patil}, {Plunkett}, {Prochaska},
  {Rastogi}, {Reddy Janga}, {Sabater}, {Sakurikar}, {Seifert}, {Sherbert},
  {Sherwood-Taylor}, {Shih}, {Sick}, {Silbiger}, {Singanamalla}, {Singer},
  {Sladen}, {Sooley}, {Sornarajah}, {Streicher}, {Teuben}, {Thomas},
  {Tremblay}, {Turner}, {Terr{\'o}n}, {van Kerkwijk}, {de la Vega}, {Watkins},
  {Weaver}, {Whitmore}, {Woillez}, {Zabalza}, \& {Contributors}}]{astropy2}
{Price-Whelan}, A.~M., {Sip{\H{o}}cz}, B.~M., {G{\"u}nther}, H.~M., {et~al.}
  2018, \aj, 156, 123, \dodoi{10.3847/1538-3881/aabc4f}

\bibitem[{{Rasio} \& {Ford}(1996)}]{R+F96}
{Rasio}, F.~A., \& {Ford}, E.~B. 1996, Science, 274, 954,
  \dodoi{10.1126/science.274.5289.954}

\bibitem[{{Rey} {et~al.}(2018){Rey}, {Bouchy}, {Stalport}, {Deleuil},
  {H{\'e}brard}, {Almenara}, {Alonso}, {Barros}, {Bonomo}, {Cazalet},
  {Delisle}, {D{\'\i}az}, {Fridlund}, {Guenther}, {Guillot}, {Montagnier},
  {Moutou}, {Lovis}, {Queloz}, {Santerne}, \& {Udry}}]{corot20_rey}
{Rey}, J., {Bouchy}, F., {Stalport}, M., {et~al.} 2018, \aap, 619, A115,
  \dodoi{10.1051/0004-6361/201833180}

\bibitem[{{Ricker} {et~al.}(2015){Ricker}, {Winn}, {Vanderspek}, {Latham},
  {Bakos}, {Bean}, {Berta-Thompson}, {Brown}, {Buchhave}, {Butler}, {Butler},
  {Chaplin}, {Charbonneau}, {Christensen-Dalsgaard}, {Clampin}, {Deming},
  {Doty}, {De Lee}, {Dressing}, {Dunham}, {Endl}, {Fressin}, {Ge}, {Henning},
  {Holman}, {Howard}, {Ida}, {Jenkins}, {Jernigan}, {Johnson}, {Kaltenegger},
  {Kawai}, {Kjeldsen}, {Laughlin}, {Levine}, {Lin}, {Lissauer}, {MacQueen},
  {Marcy}, {McCullough}, {Morton}, {Narita}, {Paegert}, {Palle}, {Pepe},
  {Pepper}, {Quirrenbach}, {Rinehart}, {Sasselov}, {Sato}, {Seager},
  {Sozzetti}, {Stassun}, {Sullivan}, {Szentgyorgyi}, {Torres}, {Udry}, \&
  {Villasenor}}]{TESS}
{Ricker}, G.~R., {Winn}, J.~N., {Vanderspek}, R., {et~al.} 2015, Journal of
  Astronomical Telescopes, Instruments, and Systems, 1, 014003,
  \dodoi{10.1117/1.JATIS.1.1.014003}

\bibitem[{{Sanchis-Ojeda} {et~al.}(2011){Sanchis-Ojeda}, {Winn}, {Holman},
  {Carter}, {Osip}, \& {Fuentes}}]{spots_wasp4}
{Sanchis-Ojeda}, R., {Winn}, J.~N., {Holman}, M.~J., {et~al.} 2011, \apj, 733,
  127, \dodoi{10.1088/0004-637X/733/2/127}

\bibitem[{Schlafly \& Finkbeiner(2011)}]{Schlafly2011}
Schlafly, E.~F., \& Finkbeiner, D.~P. 2011, Astrophysical Journal, 737,
  \dodoi{10.1088/0004-637X/737/2/103}

\bibitem[{{Schlegel} {et~al.}(1998){Schlegel}, {Finkbeiner}, \&
  {Davis}}]{Schlegel1998}
{Schlegel}, D.~J., {Finkbeiner}, D.~P., \& {Davis}, M. 1998, \apj, 500, 525,
  \dodoi{10.1086/305772}

\bibitem[{{Smith} {et~al.}(2017){Smith}, {Gandolfi}, {Barrag{\'a}n}, {Bowler},
  {Csizmadia}, {Endl}, {Fridlund}, {Grziwa}, {Guenther}, {Hatzes}, {Nowak},
  {Albrecht}, {Alonso}, {Cabrera}, {Cochran}, {Deeg}, {Cusano},
  {Eigm{\"u}ller}, {Erikson}, {Hidalgo}, {Hirano}, {Johnson}, {Korth}, {Mann},
  {Narita}, {Nespral}, {Palle}, {P{\"a}tzold}, {Prieto-Arranz}, {Rauer},
  {Ribas}, {Tingley}, \& {Wolthoff}}]{K299}
{Smith}, A.~M.~S., {Gandolfi}, D., {Barrag{\'a}n}, O., {et~al.} 2017, \mnras,
  464, 2708, \dodoi{10.1093/mnras/stw2487}

\bibitem[{{Smith} {et~al.}(2018){Smith}, {Cabrera}, {Csizmadia}, {Dai},
  {Gandolfi}, {Hirano}, {Winn}, {Albrecht}, {Alonso}, {Antoniciello},
  {Barrag{\'a}n}, {Deeg}, {Eigm{\"u}ller}, {Endl}, {Erikson}, {Fridlund},
  {Fukui}, {Grziwa}, {Guenther}, {Hatzes}, {Hidalgo}, {Howard}, {Isaacson},
  {Korth}, {Kuzuhara}, {Livingston}, {Narita}, {Nespral}, {Nowak}, {Palle},
  {P{\"a}tzold}, {Persson}, {Petigura}, {Prieto-Arranz}, {Rauer}, {Ribas}, \&
  {Van Eylen}}]{k2-137}
{Smith}, A.~M.~S., {Cabrera}, J., {Csizmadia}, S., {et~al.} 2018, \mnras, 474,
  5523, \dodoi{10.1093/mnras/stx2891}

\bibitem[{{Smith} {et~al.}(2021){Smith}, {Acton}, {Anderson}, {Armstrong},
  {Bayliss}, {Belardi}, {Bouchy}, {Brahm}, {Briegal}, {Bryant}, {Burleigh},
  {Cabrera}, {Chaushev}, {Cooke}, {Costes}, {Csizmadia}, {Eigm{\"u}ller},
  {Erikson}, {Gill}, {Gillen}, {Goad}, {G{\"u}nther}, {Henderson}, {Hogan},
  {Jord{\'a}n}, {Lendl}, {McCormac}, {Moyano}, {Nielsen}, {Rauer}, {Raynard},
  {Tilbrook}, {Turner}, {Udry}, {Vines}, {Watson}, {West}, \&
  {Wheatley}}]{ngts14}
{Smith}, A.~M.~S., {Acton}, J.~S., {Anderson}, D.~R., {et~al.} 2021, \aap, 646,
  A183, \dodoi{10.1051/0004-6361/202039712}

\bibitem[{{Smith} {et~al.}(2022){Smith}, {Breton}, {Csizmadia}, {Dai},
  {Gandolfi}, {Garc{\'\i}a}, {Howard}, {Isaacson}, {Korth}, {Lam}, {Mathur},
  {Nowak}, {P{\'e}rez Hern{\'a}ndez}, {Persson}, {Albrecht}, {Barrag{\'a}n},
  {Cabrera}, {Cochran}, {Deeg}, {Fridlund}, {Georgieva}, {Goffo}, {Guenther},
  {Hatzes}, {Kabath}, {Livingston}, {Luque}, {Palle}, {Redfield}, {Rodler},
  {Serrano}, \& {Van Eylen}}]{k299_2}
{Smith}, A.~M.~S., {Breton}, S.~N., {Csizmadia}, S., {et~al.} 2022, \mnras,
  510, 5035, \dodoi{10.1093/mnras/stab3497}

\bibitem[{{Southworth}(2011)}]{SWorth_homo4}
{Southworth}, J. 2011, \mnras, 417, 2166,
  \dodoi{10.1111/j.1365-2966.2011.19399.x}

\bibitem[{{Tinetti} {et~al.}(2018){Tinetti}, {Drossart}, {Eccleston},
  {Hartogh}, {Heske}, {Leconte}, {Micela}, {Ollivier}, {Pilbratt}, {Puig}, \&
  et~al.}]{Ariel}
{Tinetti}, G., {Drossart}, P., {Eccleston}, P., {et~al.} 2018, Experimental
  Astronomy, 46, 135, \dodoi{10.1007/s10686-018-9598-x}

\bibitem[{{Vines} \& {Jenkins}(2022)}]{ariadne}
{Vines}, J.~I., \& {Jenkins}, J.~S. 2022, \mnras, \dodoi{10.1093/mnras/stac956}

\bibitem[{{Weidenschilling} \& {Marzari}(1996)}]{W+M96}
{Weidenschilling}, S.~J., \& {Marzari}, F. 1996, \nat, 384, 619,
  \dodoi{10.1038/384619a0}

\end{thebibliography}
\bibliographystyle{aasjournal}



\end{document}